\documentclass[11pt,journal,epsfig,onecolumn,peerreviewca,draftclsnofoot]{IEEEtran}

\usepackage[dvips]{graphicx}
\usepackage{graphicx}
\usepackage{amssymb}
\usepackage{cite}
\usepackage{subfigure}
\usepackage{amsmath}
\usepackage{algorithm}
\usepackage{algorithmic}
\usepackage{multirow}

\begin{document}


\title{Compressed CPD-Based Channel Estimation and Joint Beamforming
for RIS-Assisted Millimeter Wave Communications}

\author{Xi Zheng, Jun Fang, Hongwei Wang, Peilan Wang, and Hongbin Li, ~\IEEEmembership{Fellow,~IEEE}
\thanks{Xi Zheng, Jun Fang, Hongwei Wang, and Peilan Wang are with the National Key Laboratory
of Science and Technology on Communications, University of
Electronic Science and Technology of China, Chengdu 611731, China,
Email: JunFang@uestc.edu.cn}
\thanks{Hongbin Li is
with the Department of Electrical and Computer Engineering,
Stevens Institute of Technology, Hoboken, NJ 07030, USA, E-mail:
Hongbin.Li@stevens.edu}
\thanks{This work was supported in part by the National Science
Foundation of China under Grant 61829103.}}

\maketitle


\begin{abstract}
We consider the problem of channel estimation and joint active and
passive beamforming for reconfigurable intelligent surface (RIS)
assisted millimeter wave (mmWave) multiple-input multiple-output
(MIMO) orthogonal frequency division multiplexing (OFDM) systems.
We show that, with a well-designed frame-based training protocol,
the received pilot signal can be organized into a low-rank
third-order tensor that admits a canonical polyadic decomposition
(CPD). Based on this observation, we propose two CPD-based methods
for estimating the cascade channels associated with different
subcarriers. The proposed methods exploit the intrinsic
low-rankness of the CPD formulation, which is a result of the
sparse scattering characteristics of mmWave channels, and thus
have the potential to achieve a significant training overhead
reduction. Specifically, our analysis shows that the proposed
methods have a sample complexity that scales quadratically with
the sparsity of the cascade channel. Also, by utilizing the
singular value decomposition-like structure of the effective
channel, this paper develops a joint active and passive
beamforming method based on the estimated cascade channels.
Simulation results show that the proposed CPD-based channel
estimation methods attain mean square errors that are close to the
Cram\'{e}r-Rao bound (CRB) and present a clear advantage over the
compressed sensing-based method. In addition, the proposed joint
beamforming method can effectively utilize the estimated channel
parameters to achieve superior beamforming performance.
\end{abstract}

\begin{keywords}
Reconfigurable intelligent surface, millimeter wave
communications, channel estimation, joint active and passive
beamforming.
\end{keywords}





\section{Introduction}
Millimeter wave (mmWave) and terahertz (THz) communications are
able to support extremely high data rate transmissions
\cite{RanganRappaport14,SunRappaort18}. Nevertheless, due to the
reduced diffraction effect and high penetration loss, mmWave/THz
systems require more antennas and active nodes such as access
points (APs) and relays to improve the signal coverage
\cite{AlkhateebMo14,SwindlehurstAyanoglu14}. Deploying active
nodes incurs an additional energy consumption and meanwhile
presents a serious network interference issue. It is thus of
practical significance to develop innovative technologies to
address the coverage issue of future mmWave/THz wireless networks
with a low cost and complexity.


Recently, reconfigurable intelligent surface (RIS) has emerged as
an energy-efficient and cost-effective solution to tackle the
above challenges. Generally speaking, RIS intelligently adjusts
the signal reflection through a large number of low-cost passive
reflection elements, which can dynamically reshape the wireless
propagation environment and thereby improve the system performance
\cite{LiaskosNie18,TanSun18,WuZhang19,WangFang19}. An important
advantage of RIS is that it does not require any active circuits
such as ratio frequency (RF) chains for signal
transmission/reception, which reduces hardware complexity as well
as energy consumptions compared to traditional active
transceivers/relays. Furthermore, RIS can be easily attached to
different objects (such as walls and ceilings), thus showing great
flexibility and compatibility in practical deployment
\cite{WuZhang21}.


Channel state information (CSI) acquisition is a pre-requisite to
realize the full potential of RIS-assisted mmWave systems.
Nevertheless, since RIS is usually composed of a large number of
passive elements, CSI acquisition for RIS-assisted mmWave systems
faces the difficulty of requiring a large amount of training
overhead. In \cite{YouZheng20,NingChen21,WangZhang21,WangFang22},
by utilizing the limited scattering nature of mmWave channels,
fast beam training and alignment methods were proposed for
RIS-assisted mmWave systems, where the objective is to
simultaneously identify the best beam alignment for both the base
station (BS)-RIS link and the RIS-user link. Analyses and
experimental results \cite{WangFang22} show that such methods only
require a modest amount of training overhead to establish an
effective virtual line-of-sight (LOS) path for data transmission.
Beam training methods, however, can only acquire the angular
parameters associated with the dominant path, which prevents from
utilizing the spatial diversity and achieving a higher spectral
efficiency. Different from beam training methods, other studies,
e.g. \cite{MishraJohansson19,HeYuan19,
WangFang20compressed,LiuGao20,WeiShen21,WanGao20,LinJin21,LiHuang21,AraujoAlmeida21,ZhengWang22},
aim to obtain the full CSI that suffices for active and passive
optimization. Specifically, in order to reduce the training
overhead, the inherent sparse structure of the cascade BS-RIS-user
mmWave channel was exploited and the cascade channel estimation is
cast into a compressed sensing framework
\cite{WangFang20compressed,LiuGao20,WeiShen21,WanGao20}. In
addition to compressed sensing-based methods, tensor
decomposition-based methods were developed by exploiting the
intrinsic multi-dimensional structure of the channel
\cite{LinJin21,LiHuang21,AraujoAlmeida21,ZhengWang22}. Most of
these tensor decomposition-based methods
\cite{LinJin21,LiHuang21,AraujoAlmeida21}, however, did not
utilize the sparse scattering characteristics of mmWave channels
and have a CP rank that is equal to the number of reflecting
elements. As a result, these methods require a training overhead
proportional to the number of reflecting elements, which is
usually large in practice.



In this paper, we study the problem of channel estimation and
joint beamforming for RIS-assisted mmWave MIMO-OFDM systems. We
show that by exploring the sparse scattering characteristics and
the intrinsic multi-dimensional structure of the mmWave cascade
channel, the received signal can be formulated into a low-rank
tensor that admits a canonical polyadic decomposition (CPD). Based
on this formulation, an alternating least squares (ALS) method and
a Vandemonde structured-based method are developed for channel
estimation. Theoretical analysis shows that the proposed methods
have a sample complexity of $\mathcal{O}(U^2)$. Here $U$ denotes
the sparsity of the cascade channel. Since $U$ is usually small
relative to the dimension of the cascade channel, the proposed
methods can achieve a substantial training overhead reduction. The
proposed methods, unlike compressed sensing-based techniques, are
essentially a gridless approach and therefore are free of the grid
discretization errors.


In addition to channel estimation, this paper also considers the
problem of joint active and passive beamforming design. For
point-to-point RIS-assisted MIMO systems, existing beamforming
methods, e.g. \cite{NingChen20,ZhangZhang20}, usually require the
global CSI knowledge, i.e. the channels of both the BS-RIS link
and the RIS-user link, for joint beamforming. In this paper, by
exploring the inherent structure of the effective channel, we show
that the knowledge of the cascade channel alone suffices for the
purpose of joint active and passive beamforming, and we develop a
manifold optimization-based scheme for active and passive
optimization based on the estimated cascade channel.


The current work is an extension of our previous work
\cite{ZhengWang22} which developed a CPD-based channel estimation
method for RIS-assisted mmWave MISO systems. The extension of the
current work consists of two aspects. First, we extend the
CPD-based method to the MIMO scenarios. Note that such an
extension is nontrivial. In fact, for the MISO case, as pointed
out in \cite{ZhengWang22}, the Kruskal's condition which is
essential to the uniqueness of the CPD does not hold and hence the
classical ALS method cannot be applied. We show that due to the
diversity brought by multi-antenna at the receiver, the Kruskal's
condition can be satisfied for MIMO scenarios. Such a fact enables
us to develop an ALS-based channel estimation method that is more
robust against noise. Second, besides channel estimation, the
current work also considers how to optimize the active and passive
beamforming coefficients based on the estimated cascade channel,
which is a challenging problem (particularly for MIMO systems) and
was not studied in our previous work.


The rest of the paper is organized as follows. In Section
\ref{sec:system-model}, the system model and the formulation of
the channel estimation problem are discussed. CPD-based channel
estimation methods are developed in Section \ref{sec:CPD-method}.
In Section \ref{sec:beamforming}, the problem of joint active and
passive beamforming design is studied. Finally, simulation results
are presented in Section \ref{sec:results}.

\begin{figure}[t]
\centering
\includegraphics[width=8cm]{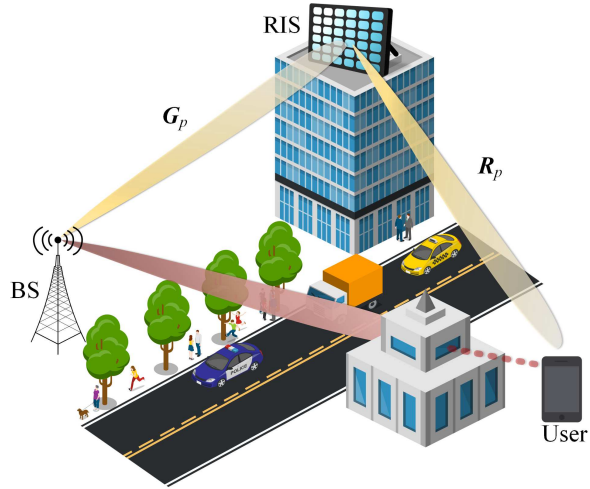}
\caption{RIS-assisted mmWave MIMO-OFDM systems.}
\label{fig1}
\end{figure}

\section{System Model and Problem Formulation} \label{sec:system-model}
Consider a point-to-point RIS-aided mmWave MIMO-OFDM system, where
an RIS is deployed to assist the downlink data transmission of
$N_{\rm s}$ data streams from the BS to the user (see Fig.
\ref{fig1}). For simplicity, we assume that the direct link
between the BS and the user is blocked due to poor propagation
conditions. The BS is equipped with a uniform linear array (ULA)
with $N_{\rm t}$ antennas and $R_{\rm t}$ radio frequency (RF)
chains, and the user is equipped with a ULA with $N_{\rm r}$
antennas and $R_{\rm r}$ RF chains, where $R_{\rm t}\ll N_{\rm t}$
and $R_{\rm r}\ll N_{\rm r}$. The RIS is a uniform planar array
(UPA) with $M = M_{\rm y} \times M_{\rm z}$ passive reflecting
elements. Each element, say the $m$th element, can independently
reflect the incident signal with a reconfigurable phase shift
$e^{j\varsigma_{m}}$. For notational simplicity, let
$\boldsymbol{v}\triangleq
[e^{j\varsigma_{1}}\phantom{0}\ldots\phantom{0}e^{j\varsigma_{M}}]^H$
denote the reflection coefficient vector, and
$\mathbf{\Phi}\triangleq\text{diag}(\boldsymbol{v}^H)$ denote the
reflection matrix.

\begin{figure*}[ht]
\centering
\includegraphics[width=16cm]{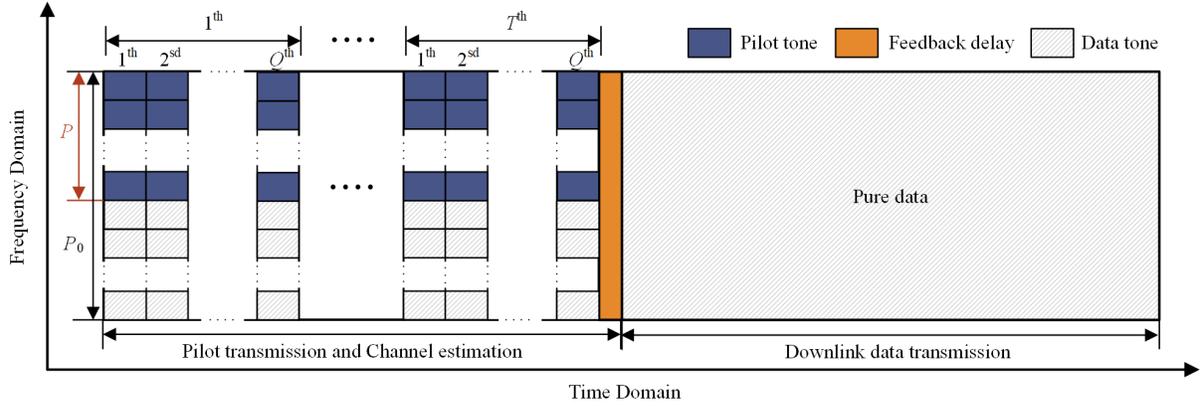}
\caption{Illustration of the proposed transmission protocol.}
\label{fig2}
\end{figure*}

\subsection{Channel Model}
In this paper, we adopt a geometric wideband mmWave channel model
\cite{AlkhateebHeath16} to characterize the channel between the BS
(RIS) and the IRS (user). Specifically, the BS-RIS channel in the
delay domain can be expressed as
\begin{equation}
\boldsymbol{G}(\tau)=\sum_{l=1}^{L} \alpha_{l}
\boldsymbol{a}_{\mathrm{IRS}} \left(\vartheta_{l}^{{\rm r}},
\chi_{l}^{{\rm r}}\right) \boldsymbol{a}_{\mathrm{BS}}^{H}
\left(\phi_{l}\right) \delta\left(\tau-\tau_{l}\right) \\
\end{equation}
where $L$ is the total number of paths between the BS and the RIS,
$\alpha_l$ is the complex gain associated with the $l$th path,
$\phi_{l}$ represents the spatial angle of departure (AoD),
$\{\vartheta_{l}^{{\rm r}}, \chi_{l}^{{\rm r}}\}$ denote the
spatial azimuth and elevation angles of arrival (AoAs), $\tau _l$
denotes the time delay, $\delta \left( \tau \right)$ denotes the
Dirac-delta function,
$\boldsymbol{a}_{{\rm{IRS}}}(\vartheta,\chi)$ and
$\boldsymbol{a}_{\rm{BS}}(\phi)$ denote the normalized array
response vectors associated with the IRS and the BS, respectively.
Similarly, the IRS-user channel in the delay domain is modeled as
\begin{equation}
\boldsymbol{R}(\tau)=\sum_{l=1}^{L_{{\rm r}}} \beta_{l}
{{\boldsymbol{a}}_{\rm{UE}}}\left( {{\theta }_{l}}
\right)\boldsymbol{a}_{\rm{IRS}}^{H}\left( \vartheta _{l}^{{\rm
t}},\chi _{l}^{{\rm t}} \right) \delta\left(\tau-\kappa_{l}\right)
\end{equation}
where $L_{\rm r}$ is the number of paths between the RIS and the
user, $\beta_{l}$ denotes the associated complex path gain,
$\theta_{l}$ represents the spatial AoA, $\{\vartheta_{l}^{{\rm
t}}, \chi_{l}^{\rm t}\}$ denote the spatial azimuth and elevation
AoDs, $\kappa_{l}$ is the time delay, and
$\boldsymbol{a}_{\rm{UE}}(\theta)$ denotes the normalized array
response vectors associated with the user. The normalized array
response vectors $\boldsymbol{a}_{\rm{BS}}(\phi)$,
$\boldsymbol{a}_{\rm{UE}}(\theta)$ and
$\boldsymbol{a}_{{\rm{IRS}}}(\vartheta,\chi)$ are respectively
defined as
\begin{align}
\boldsymbol{a}_{\mathrm{BS}}(\phi) \triangleq
\frac{1}{\sqrt{N_{\rm t}}}[ 1 \phantom{0} e^{j \phi} \phantom{0}
\cdots \phantom{0} e^{j(N_{\rm t}-1)\phi} ]^{T}
\end{align}
\begin{align}
\boldsymbol{a}_{\mathrm{UE}}(\theta) \triangleq
\frac{1}{\sqrt{N_{\rm r}}}[1 \phantom{0} e^{j \theta} \phantom{0}
\cdots \phantom{0} e^{j(N_{\rm r}-1) \theta} ]^{T}
\end{align}
\begin{align}
\boldsymbol{a}_{\rm{IRS}}(\vartheta,\chi)& \triangleq
\boldsymbol{a}_{\rm y}(\vartheta) \otimes \boldsymbol{a}_{\rm
z}(\chi)  \\
&=\frac{1}{\sqrt{M_{\rm y}}}[1 \phantom{0} e^{j \vartheta}
\phantom{0} \cdots \phantom{0} e^{j(M_{\rm y}-1)\vartheta}]^{T} \\
&\ \ \otimes \frac{1}{\sqrt{M_{\rm z}}}[1 \phantom{0} e^{j \chi}
\phantom{0} \cdots \phantom{0} e^{j(M_{\rm z}-1) \chi} ]^{T}
\end{align}
where $\otimes$ denotes the Kronecker product, $\phi
\triangleq\frac{2\pi d}{\lambda}\sin(\eta)$, $\theta\triangleq
\frac{2\pi d}{\lambda}\sin(\gamma)$, $\vartheta\triangleq
\cos(\varpi)\sin(\psi)$, and $\chi\triangleq\sin(\varpi)$. Here
$\psi$ and $\varpi$ denote the azimuth and elevation angles
associated with the IRS, $\gamma$ ($\eta$) represents the angle
associated with the user (BS), $d$ and $\lambda$ denote the
adjacent spacing and the signal wavelength, respectively.

Accordingly, the frequency-domain BS-RIS and RIS-user channels
associated with the $p$th subcarrier can be respectively expressed
as
\begin{equation}
{{\boldsymbol{G}}_p} = \sum\limits_{l = 1}^L {{\alpha _l}} {e^{ -
j2\pi {f_{\rm s}}{\tau
_l}\frac{p}{{{P_0}}}}}{{\boldsymbol{a}}_{{\rm{IRS}}}}\left(
{\vartheta _l^{\rm r},\chi _l^{\rm r}}
\right){\boldsymbol{a}}_{{\rm{BS}}}^H\left( {{\phi _l}} \right)
\label{Gp}
\end{equation}
\begin{equation}
{{\boldsymbol{R}}_p} = \sum\limits_{l = 1}^{{L_{\rm r}}} {{\beta
_l}} {e^{ - j2\pi {f_{\rm s}}{\kappa
_l}\frac{p}{{{P_0}}}}}{{\boldsymbol{a}}_{{\rm{UE}}}}\left(
{{\theta _l}} \right){\boldsymbol{a}}_{{\rm{IRS}}}^H\left(
{\vartheta _l^{\rm t},\chi _l^{\rm t}} \right) \label{Rp}
\end{equation}
where $f_{\rm s}$ is the sample frequency, and $P_0$ denotes the
total number of OFDM tones.


\subsection{Downlink Training and Signal Model}
To facilitate the algorithm development, we employ a frame-based
downlink training protocol (see Fig. \ref{fig2}). Specifically,
the training period is divided into $T$ time frames, where the BS
(user) employs different beamforming (combining) vectors at
different time frames. Each time frame is further divided into $Q$
time slots. At the $q$th time slot, the RIS adopts an individual
phase-shift matrix $\mathbf{\Phi}_q$ to reflect the impinging
signal. Suppose the total number of OFDM tones is $P_0$, among
which $P$, say $\{1,2,...,P\}$, subcarriers are selected for
training. The transmitted signal associated with the $p$th
subcarrier at the $t$th time frame can be expressed as
\begin{align}
\boldsymbol{f}_{t,p}=\boldsymbol{F}_{\text{RF},t}\boldsymbol{F}_{\text{BB},t,p}{\boldsymbol{s}_{t,p}}\in
\mathbb{C}^{N_{\rm t}}
\end{align}
where $\boldsymbol{s}_{t,p} \in \mathbb{C}^{N_{\rm s}}$ denotes
the $p$th subcarrier's pilot symbol vector,
$\boldsymbol{F}_{\text{BB},t,p}\in \mathbb{C}^{R_{\rm t} \times
N_{\rm s}}$ is the baseband precoding matrix associated with the
$p$th subcarrier, and
$\boldsymbol{F}_{\text{RF},t}\in\mathbb{C}^{N_{\rm t}\times R_{\rm
t}}$ is a radio frequency (RF) precoder common to all subcarriers.
The transmitted signal arrives at the user via propagation through
the BS-RIS-user channel. The user employs the hybrid combiner
${{\boldsymbol{W}}}_{t,p} = {\boldsymbol{W}_{{\rm
RF},t}}{\boldsymbol{W}_{{\rm BB},t,p}}\in \mathbb{C}^{N_{\rm r}
\times N_{\rm s}}$ to combine the received signal, where
${\boldsymbol{W}_{{\rm BB},t,p}}\in \mathbb{C}^{R_{\rm r} \times
N_{\rm s}}$ denotes the baseband combiner associated with the
$p$th subcarrier and $\boldsymbol{W}_{{\rm RF},t} \in
\mathbb{C}^{N_{\rm r} \times R_{\rm r}}$ is a RF combiner common
to all subcarriers. Hence, the received signal
${\boldsymbol{y}_{q,t,p}}\in \mathbb{C}^{N_{\rm s}}$ associated
with the $p$th sub-carrier at the $q$th time slot of the $t$th
time frame can be written as
\begin{align}
\label{y_transmit}
\boldsymbol{y}_{q,t,p}=\boldsymbol{W}^{H}_{\mathrm{BB}, t,p}
\boldsymbol{W}^{H}_{\mathrm{RF},t}
{{\boldsymbol{R}}_p}{{\boldsymbol{\Phi
}}_q}{{\boldsymbol{G}}_p}\boldsymbol{F}_{\mathrm{RF},t}
\boldsymbol{F}_{\mathrm{BB},t,p} {\boldsymbol{s}_{t
,p}}+{\boldsymbol{n}_{q,t,p}}
\end{align}
In the channel estimation stage, for simplicity, we assume
$\boldsymbol{F}_{\text{BB},t,p}=\boldsymbol{F}_{\text{BB},t}$ and
$\boldsymbol{s}_{t,p}=\boldsymbol{s}_t$, in which case we have
$\boldsymbol{f}_{t,p}={\boldsymbol{f}_{t}}\triangleq
\boldsymbol{F}_{\text{RF},t}\boldsymbol{F}_{\text{BB},t}{\boldsymbol{s}_t}
$. Similarly, let ${\boldsymbol{W}_{{\rm
BB},t,p}}={\boldsymbol{W}_{{\rm BB},t}}$. We have
${{\boldsymbol{W}}}_{t,p} = \boldsymbol{W}_{t}
\triangleq{\boldsymbol{W}_{{\rm RF},t}}{\boldsymbol{W}_{{\rm
BB},t}}$. The received signal can thus be expressed as
\begin{align}
{\boldsymbol{y}_{q,t,p}} &= {\boldsymbol{W}_t^H}{{\boldsymbol{R}}_p}{{\boldsymbol{\Phi }}_q}
{{\boldsymbol{G}}_p}{\boldsymbol{f}_t} + {\boldsymbol{n}_{q,t,p}} \nonumber\\
 &=(\boldsymbol{f}_t^T\otimes\boldsymbol{W}_t^H){\rm{vec}}
 ({\boldsymbol{R}}_p{{\boldsymbol{\Phi }}_q}{{\boldsymbol{G}}_p}) + {\boldsymbol{n}_{q,t,p}} \nonumber\\
 &={{\boldsymbol{X}_t^T}}({{\boldsymbol{G}}^T_p}\odot {\boldsymbol{R}}_p)
 \boldsymbol{v}_q^* + {\boldsymbol{n}_{q,t,p}}
\label{y_channel}
\end{align}
where we define
${\boldsymbol{X}_t}\triangleq\boldsymbol{f}_t\otimes\boldsymbol{W}^*_t
\in \mathbb{C}^{{N_{\rm t}}{N_{\rm r}}\times {N_{\rm s}}}$,
$\odot$ denotes the Khatri-Rao product, and $\boldsymbol{H}_p
\triangleq {\boldsymbol{G}}_p^T\odot
{\boldsymbol{R}}_p\in\mathbb{C}^{N_{\rm t}N_{\rm r}\times M}$
denotes the cascade channel associated with the $p$th subcarrier.
Recalling (\ref{Gp}) and (\ref{Rp}), the cascade channel
$\boldsymbol{H}_p$ can be further expressed as
\begin{align}
\boldsymbol{H}_{p}&=\boldsymbol{G}_{p}^{T} \odot \boldsymbol{R}_{p} \nonumber\\
&=\left(\sum_{m=1}^{L} \alpha_{m} e^{-j 2 \pi f_{\rm s} \tau_{m}
\frac{p}{P_{0}}} \boldsymbol{a}^{*}_{\rm{BS}}\left(\phi_{m}\right)
\boldsymbol{a}_{\rm{IRS}}^{T}\left(\vartheta_{m}^{\rm r},
\chi_{m}^{\rm r}\right)\right) \nonumber\\
&\ \ \ \ \odot\left(\sum_{n=1}^{L_{\rm r}} \beta_{n} e^{-j 2 \pi
f_{\rm s} \kappa_{n} \frac{p}{P_{0}}}
\boldsymbol{a}_{\rm{UE}}\left(\theta_{n}\right)
\boldsymbol{a}_{\rm{IRS}}^{H}\left(\vartheta_{n}^{\rm t}, \chi_{n}^{\rm t}\right)\right) \nonumber\\
&=\sum_{m=1}^{L} \sum_{n=1}^{L_{\rm r}} \alpha_{m} \beta_{n} e^{-j
2 \pi f_{\rm s}\left(\tau_{m}+\kappa_{n}\right) \frac{p}{p_{0}}}
\times \left(\boldsymbol{a}^{*}_{\rm{BS}}\left(\phi_{m}\right)
\otimes \boldsymbol{a}_{\rm{UE}}
\left(\theta_{n}\right)\right)\nonumber\\
&\ \ \ \ \ \ \ \ \ \ \ \ \ \ \ \ \ \ \ \ \ \times
\left(\boldsymbol{a}_{\rm{IRS}}^{T}\left(\vartheta_{m}^{\rm r}, \chi_{m}^{\rm r}\right)
\odot \boldsymbol{a}_{\rm{IRS}}^{H}\left(\vartheta_{n}^{\rm t}, \chi_{n}^{\rm t}\right)\right)\nonumber\\
&=\sum_{m=1}^{L} \sum_{n=1}^{L_{\rm r}} \alpha_{m} \beta_{n} e^{-j 2 \pi f_{\rm s}
\left(\tau_{m}+\kappa_{n}\right) \frac{p}{p_{0}}}{\boldsymbol{a}}_{{\rm{S}}}\left( \phi_{m},\theta _{n} \right)
\nonumber\\
&\ \ \ \ \ \ \ \ \ \ \ \ \ \ \ \ \ \ \ \ \ \ \ \times
\boldsymbol{a}_{\rm{IRS}}^{T}\left(\vartheta_{m}^{\rm
r}-\vartheta_{n}^{\rm t},
\chi_{m}^{\rm r}-\chi_{n}^{\rm t}\right)\nonumber\\
&\mathop  = \limits^{(a)} \sum\limits_{u = 1}^{L{L_{\rm r}}}
{{\varrho _u}} {e^{ - j2\pi {f_{\rm s}}{\iota
_u}\frac{p}{{{P_0}}}}}{\boldsymbol{a}}_{{\rm{S}}}\left(
\phi_{u},\theta _{u} \right){\boldsymbol{a}_{\rm{IRS}}^T}\left(
{{\zeta _u},{\xi _u}} \right)\label{H}
\end{align}
where ${\boldsymbol{a}}_{{\rm{S}}}\left( \phi_{m},\theta _{n}
\right)\triangleq
\boldsymbol{a}^{*}_{\rm{BS}}\left(\phi_{m}\right) \otimes
\boldsymbol{a}_{\rm{UE}}\left(\theta_{n}\right)$, and the mapping
process $\left(a\right)$ is defined as
\begin{gather}
\nonumber {(m - 1)L_{\rm r} + n \mapsto u,u = 1, \ldots ,L{L_{\rm
r }}}\nonumber\\ {{\alpha _m}{\beta _n} \mapsto {\varrho _u},u =
1, \ldots ,L{L_{\rm r}}}\nonumber\\
{{\tau _m} + {\kappa _n} \mapsto {\iota _u},u = 1, \ldots ,L{L_{\rm r}}}\nonumber\\
\vartheta _m^{\rm r} - \vartheta _n^{\rm t} \mapsto \zeta _u,u =
1, \ldots ,L{L_{\rm r}}\nonumber\\ \chi _m^{\rm r} - \chi _n^{\rm
t} \mapsto \xi _u,u = 1, \ldots ,L{L_{\rm r}}\nonumber \\ \phi _m
\mapsto \phi _u,m = \left\lceil {\frac{u}{L_{\rm r}}} \right\rceil
,u = 1, \ldots ,L{L_{\rm r}}\nonumber \\ \theta _n \mapsto \theta
_u,n = u-\left(\left\lceil {\frac{u}{L_{\rm r}}}
\right\rceil-1\right)L_{\rm r},u = 1, \ldots ,L{L_{\rm r}}
\label{mapping}
\end{gather}
where $\left\lceil x \right\rceil$ denotes the ceiling function
which gives the least integer greater than or equal to $x$.

Our objective of this paper is two-fold. First, we wish to develop
a method to estimate the cascade channel matrices
$\{\boldsymbol{H}_{p}\}$ from the received measurements
$\{\boldsymbol{y}_{q,t,p}\}$. In particular, the proposed method
is expected to provide a reliable channel estimate by using as few
pilot symbols as possible. After the cascade channel matrices are
obtained, another purpose of this work is to develop a joint
active and passive beamforming method which aims to maximize the
spectral efficiency by exploiting the knowledge of the cascade
channels.

\section{Proposed CPD-Based Channel Estimation Method}
\label{sec:CPD-method}
\subsection{Low-Rank Tensor Representation}
Substituting (\ref{H}) into (\ref{y_channel}), we have
\begin{align}
{\boldsymbol{y}_{q,t,p}}
&= \sum\limits_{u = 1}^{L{L_{\rm r}}} {{\varrho _u}}
{e^{ - j2\pi {f_{\rm s}}{\iota _u}\frac{p}{{{P_0}}}}}{{\boldsymbol{X}_t^T}}{\boldsymbol{a}}_{{\rm{S}}}
\left( \phi_{u},\theta _{u} \right) \nonumber \\
&\ \ \ \ \ \ \ \ \ \ \ \ \ \ \ \ \ \ \ \ \ \ \ \ \ \times
{\boldsymbol{a}_{\rm{IRS}}^T}\left( {{\zeta _u},{\xi _u}}
\right)\boldsymbol{v}_q^*  + {\boldsymbol{n}_{q,t,p}}
\end{align}
Define $\boldsymbol{Y}_{t,p} \triangleq
\left[\boldsymbol{y}_{1,t,p}\phantom{0} \cdots\phantom{0}
\boldsymbol{y}_{Q,t,p}\right]^T \in \mathbb{C}^{Q\times N_{\rm
s}}$. The received signal at the $t$th time frame can be written
as
\begin{align}
{\boldsymbol{Y}_{t,p}}
&= \sum\limits_{u = 1}^{L{L_{\rm r}}} {{\varrho _u}} {e^{ - j2\pi {f_{\rm s}}{\iota _u}
\frac{p}{{{P_0}}}}}\boldsymbol{V}^T{\boldsymbol{a}_{\rm{IRS}}}\left( {{\zeta _u},{\xi _u}} \right)
\nonumber\\
&\ \ \ \ \ \ \ \ \ \ \ \ \ \ \ \ \ \ \ \ \ \ \ \ \ \times
{\boldsymbol{a}}^T_{{\rm{S}}}\left( \phi_{u},\theta _{u}
\right){{\boldsymbol{X}_t}}  + {\boldsymbol{N}_{t,p}}
\end{align}
where
\begin{gather}
\boldsymbol{V} \triangleq \left[\boldsymbol{v}_{1}^*\phantom{0} \cdots\phantom{0}
\boldsymbol{v}_{Q}^*\right] \in \mathbb{C}^{M \times Q} \\
\boldsymbol{N}_{t,p}\triangleq\left[\boldsymbol{n}_{1,t,p}\phantom{0}
\cdots\phantom{0} \boldsymbol{n}_{Q,t,p}\right]^T \in
\mathbb{C}^{Q \times N_{\rm s}}
\end{gather}
Collecting the received signals from all time frames and defining
$\boldsymbol{Y}_p \triangleq \left[\boldsymbol{Y}_{1,p}\phantom{0}
\cdots\phantom{0}\boldsymbol{Y}_{T,p}\right] \in \mathbb{C}^{Q
\times TN_{\rm s}}$, we have
\begin{align}
{{\boldsymbol{Y}}_p} &= \sum\limits_{u = 1}^{L{L_{\rm r}}}
{{\varrho _u}} {e^{ - j2\pi {f_{\rm s}}{\iota
_u}\frac{p}{{{P_0}}}}}\boldsymbol{V}^T{\boldsymbol{a}_{\rm{IRS}}}
\left( {{\zeta _u},{\xi _u}} \right) \nonumber \\
&\ \ \ \ \ \ \ \ \ \ \ \ \ \ \ \ \ \ \ \ \ \ \ \ \ \ \ \ \ \ \ \ \ \ \ \ \times {\boldsymbol{a}}^T_{{\rm{S}}}
\left( \phi_{u},\theta _{u} \right)\boldsymbol{F}+ {{\boldsymbol{N}}_p}\nonumber\\
&= \sum\limits_{u = 1}^{L{L_{\rm r}}} {{\varrho _u}} {e^{ - j2\pi
{f_s}{\iota _u}\frac{p}{{{P_0}}}}}{{{\boldsymbol{\tilde
a}}}_{{\rm{IRS}}}}\left( {{\zeta _u},{\xi _u}}
\right){\boldsymbol{\tilde a}}^T_{\rm{S}}\left( {{\phi _u},{\theta
_u}} \right) + {{\boldsymbol{N}}_p}
\end{align}
where ${{{\boldsymbol{\tilde a}}}_{{\rm{IRS}}}}\left( {{\zeta
_u},{\xi _u}} \right) \triangleq \boldsymbol{V}^T
{\boldsymbol{a}_{\rm{IRS}}}\left( {{\zeta _u},{\xi _u}} \right)
\in \mathbb{C}^{Q}$, ${\boldsymbol{\tilde a}}_{\rm{S}}\left(
{{\phi _u},{\theta _u}} \right) \triangleq \boldsymbol{F}^T
{\boldsymbol{a}}_{{\rm{S}}}\left( \phi_{u},\theta _{u} \right) \in
\mathbb{C}^{TN_{\rm s}}$, and
\begin{gather}
{\boldsymbol{F}}  \triangleq \left[{\boldsymbol{X}_1}\phantom{0}\cdots\phantom{0}
{\boldsymbol{X}_T} \right] \in \mathbb{C}^{{N_{\rm t}}{N_{\rm r}} \times {TN_{\rm s}}}\\
{{\boldsymbol{N}}_p} \triangleq \left[
{{\boldsymbol{N}}_{1,p}\phantom{0}\cdots\phantom{0}{\boldsymbol{N}}_{T,p}}
\right] \in \mathbb{C}^{ Q\times TN_{\rm s}}
\end{gather}
Since signals associated with multiple subcarriers are available,
the entire received signal $\{\boldsymbol{Y}_p\}_p$ can be
expressed as a third-order tensor $\boldsymbol{\cal{Y}}\in
\mathbb{C}^{Q \times TN_{\rm s} \times P}$ whose three modes
respectively represent the sub-frame, the time frame and the
subcarrier. Each slice of the tensor ${\boldsymbol{\cal Y}}$,
${\boldsymbol{Y}}_p$, is a weighted sum of a common set of
rank-one outer products. Hence the third-order tensor
${\boldsymbol{\cal Y}}$ admits a CPD which decomposes a tensor
into a sum of rank-one component tensors, i.e.
\begin{equation} \label{eqn18}
{\boldsymbol{\cal Y}} = \sum\limits_{u = 1}^U
{{{{\boldsymbol{\tilde a}}}_{{\rm{IRS}}}}} \left( {{\zeta _u},{\xi
_u}} \right) \circ \left( {{\varrho _u}{\boldsymbol{\tilde
a}}_{\rm{S}}\left( {{\phi _u},{\theta _u}} \right)} \right) \circ
{\boldsymbol{g}}\left( {{\iota _u}} \right) + \boldsymbol{\cal N}
\end{equation}
where $U \triangleq LL_{\rm r}$, $\boldsymbol{\cal N}\in
\mathbb{C}^{ Q\times TN_{\rm s}\times P}$ is the noise tensor, and
\begin{equation}
{\boldsymbol{g}}\left( {{\iota _u}} \right) \triangleq [{{e^{ -
j2\pi {f_{\rm s}}{\iota _u}\frac{1}{{{P_0}}}}}\phantom{0} \cdots
\phantom{0}{e^{ - j2\pi {f_{\rm s}}{\iota _u}\frac{P}{{{P_0}}}}}}
]^T
\end{equation}
Define
\begin{align}
\boldsymbol{A} \triangleq &\left[ {{{{\boldsymbol{\tilde
a}}}_{{\rm{IRS}}}}\left( {{\zeta _1},{\xi _1}} \right)\phantom{0}
\cdots \phantom{0}{{{\boldsymbol{\tilde a}}}_{{\rm{IRS}}}}\left(
{{\zeta _U},{\xi
_U}} \right)} \right] \in \mathbb{C}^{Q\times U}\\
\boldsymbol{B} \triangleq & \left[ { {{\varrho
_1}{{{\boldsymbol{\tilde a}}}_{\rm{S}}} \left( {{\phi _1},{\theta
_1}} \right)} \phantom{0} \cdots \phantom{0} {{\varrho
_U}{{{\boldsymbol{\tilde a}}}_{\rm{S}}}
\left( {{\phi _U},{\theta _U}} \right)} } \right] \in \mathbb{C}^{TN_{\rm s} \times U}\\
\boldsymbol{C} \triangleq & \left[ {{\boldsymbol{g}}\left( {{\iota
_1}} \right)\phantom{0} \cdots \phantom{0}{\boldsymbol{g}}\left(
{{\iota _U}} \right)} \right] \in \mathbb{C}^{P \times U}
\end{align}
Here $\{\boldsymbol{A},\boldsymbol{B},\boldsymbol{C}\}$ are the
factor matrices of the tensor $\boldsymbol{\cal Y}$. We see that
the channel parameters can be readily estimated from these factor
matrices. Therefore our objective is to first obtain the factor
matrices from the observed tensor $\boldsymbol{\cal Y}$, and then
estimate the associated channel parameters from the estimated
factor matrices.


\subsection{Uniqueness Condition}
Before proceeding to the CPD, we first study the condition that
ensures the uniqueness of the CPD since the uniqueness of the CPD
is essential to the success of our proposed method. This condition
also sheds light on the sample complexity of the proposed method,
i.e. the amount of training overhead required to reliably estimate
the channel.

A well-known sufficient condition for the uniqueness of the CPD is
known as Kruskal's condition summarized as follows
\cite{KruskalJoseph77}.

\newtheorem{theorem}{Theorem}
\begin{theorem} \label{theorem1}
Let ${\boldsymbol{\cal X}}\in\mathbb{C}{^{I \times J \times K}}$
be a third-order tensor decomposed of three factor matrices
${\boldsymbol{A}^{\left(1\right)}} \in \mathbb{C}^{I \times R}$,
${\boldsymbol{A}^{\left(2\right)}} \in\mathbb{C}^ {{J \times R}}$
and ${\boldsymbol{A}^{\left(3\right)}} \in\mathbb{C} {^{K \times
R}}$. If the condition
\begin{align}
k_{\boldsymbol{A}^{\left(1\right)}} +
k_{\boldsymbol{A}^{\left(2\right)}} +
k_{\boldsymbol{A}^{\left(3\right)}} \ge 2R + 2
\end{align}
is satisfied, then the CPD of ${\boldsymbol{{\cal X }}}$ is unique
up to scaling and permutation ambiguities.
\end{theorem}

Here $k_{\boldsymbol{A}}$ denotes the k-rank of $\boldsymbol{A}$,
which is defined as the largest value of $k_{\boldsymbol{A}}$ such
that every subset of $k_{\boldsymbol{A}}$ columns of
$\boldsymbol{A}$ is linearly independent. From the above theorem,
we know that if
\begin{align}
k_{\boldsymbol{A}} + k_{\boldsymbol{B}} + k_{\boldsymbol{C}} \ge
2U + 2
\end{align}
then the CP decomposition of $\boldsymbol{\cal{Y}}$ is essentially
unique.

We first examine the k-rank of $\boldsymbol{A}$. Recall that
\begin{align}
{\boldsymbol{A}} \triangleq {{\boldsymbol{V}}^T}\left[
{{{\boldsymbol{a}}_{{\rm{IRS}}}}\left( {{\zeta _1},{\xi _1}}
\right)}\phantom{0} \cdots\phantom{0}
{{{\boldsymbol{a}}_{{\rm{IRS}}}}\left( {{\zeta _U},{\xi _U}}
\right)} \right]
\end{align}
Assume that each entry of ${\boldsymbol{V}} \in \mathbb{C}^{M
\times Q}$ is randomly generated from the unit circle, i.e.
${v_{m,n}} = {e^{j{\sigma_{m,n}}}}$, where ${\sigma_{m,n}} \in
\left[ { - \pi ,\pi } \right]$ is drawn from a uniform
distribution. Let $a_{m,i} \triangleq
{\boldsymbol{v}}_m^T{{\boldsymbol{a}}_{{\rm{IRS}}}}\left( {{\zeta
_i},{\xi _i}} \right)$ denote the $(m,i)$th entry of
$\boldsymbol{A}$, where ${\boldsymbol v_{m}}$ is the $m$th column
of $\boldsymbol V$. It can be easily verified that $E\left[
{{a_{m,i}}} \right]=0,\forall m,i$ and
\begin{align}
E\left[ {{a^{H}_{m,i}}a_{n,j} } \right] = \left\{
{\begin{array}{*{20}{l}}
0&{m \ne n}\\
{{\boldsymbol{a}}_{{\rm{IRS}}}^H\left( {{\zeta _j},{\xi _j}}
\right){{\boldsymbol{a}}_{{\rm{IRS}}}}\left( {{\zeta _i},{\xi _i}}
\right)}&{m = n}
\end{array}} \right.
\end{align}
Recall that $\boldsymbol{a}_{\rm{IRS}}(\zeta_i,\xi_i)$ is a
Kronecker product of two steering vectors. Hence we have
${\boldsymbol{a}}_{{\rm{IRS}}}^H\left( {{\zeta _j},{\xi _j}}
\right){{\boldsymbol{a}}_{{\rm{IRS}}}}\left( {{\zeta _i},{\xi _i}}
\right) \approx 0$, when ${\zeta _i} \ne {\zeta _j}$ or ${\xi _i}
\ne {\xi _j}$ \cite{chenjin13}. In reality, due to the random
nature of the channel parameters, the angles $\{\zeta_u\}_{u=1}^U$
are mutually distinct with probability one, and so are the angles
$\{\xi_u\}_{u=1}^U$. We thus have $E\left[ {{a^{H}_{m,i}}a_{n,j}}
\right]\approx 0$ even for the case $m=n$. On the other hand,
according to the central limit theorem, $a_{m,i}$ approximately
follows a Gaussian distribution. Therefore entries of
$\boldsymbol{A}$ can be considered as i.i.d. Gaussian variables
with zero mean and unit variance. As a result, we have
\begin{align}
{k_{\boldsymbol{A}}} = \min \left\{ {Q,U} \right\}
\end{align}

The factor matrix $\boldsymbol{B}$ can be written as
\begin{align}
{\boldsymbol{B}} &\triangleq {{\boldsymbol{F}}^T}\left[
{{{\boldsymbol{a}}_{\rm{S}}}\left( {{\phi _1},{\theta _1}}
\right)}\phantom{0}
\cdots\phantom{0}{{{\boldsymbol{a}}_{\rm{S}}}\left( {{\phi
_U},{\theta _U}} \right)}
 \right]\nonumber\\
&\ \ \ \ \ \ \ \ \ \ \ \ \ \ \ \ \ \ \ \ \ \ \
\times{\rm{diag}}\left( {{\varrho _1}, \cdots,{\varrho _U}}
\right)
\end{align}
Similarly, for a randomly generated $\boldsymbol{F}$ whose entries
are uniformly chosen from a unit circle, we can arrive at the
conclusion that the $k$-rank of $\boldsymbol{B}$ is equal to
\begin{align}
{k_{\boldsymbol {B}}} = \min \left\{ {TN_{\rm s},U} \right\}
\end{align}
Note that due to the mapping process, the set $\{\phi_u\}$ only
contains $L$ distinct elements, and $\{\theta_u\}$ only contains
$L_r$ distinct elements. Nevertheless, for $i\neq j$, we still
have
$\boldsymbol{a}_{\rm{S}}(\phi_i,\theta_i)\neq\boldsymbol{a}_{\rm{S}}(\phi_j,\theta_j)$
since each pair of $(\phi_u,\theta_u)$ is unique according to the
mapping rule defined in (\ref{mapping}).


As for the factor matrix ${\boldsymbol {C}}$, it is a Vandermonde
matrix with distinct generators, i.e. ${\iota _i} \ne {\iota _j},
\forall i\neq j$. Thus we have
\begin{align}
{k_{\boldsymbol{C}}} = \min \left\{ {P,U} \right\}
\end{align}

Based on the above results, we know that Kruskal's condition is
equivalent to
\begin{align}\label{uniqueness_1}
\min \left\{ {Q,U} \right\} + \min \left\{ {TN_{\rm s},U} \right\}
+ \min \left\{ {P,U} \right\} \ge 2U + 2
\end{align}
Note that the total number of pilot signals for downlink training
is $QTP$. To meet condition (\ref{uniqueness_1}), we can set
$Q\geq U$, $TN_{\rm s}\geq U$ and $P\geq 2$, in which case the
amount of training overhead is in the order of
$\mathcal{O}(2U^{2}/N_{s})$. We see that the sample complexity of
the proposed method only depends on the sparsity of the cascade
channel $U$. As $U$ is usually small relative to the dimension of
the cascade channel, a substantial training overhead reduction can
be achieved.

\subsection{CPD: An ALS-Based Approach}
For generic CPD problems, an alternating least squares (ALS)
method is usually employed to search for the factor matrices.
Specifically, assume that the CP rank, $U$, is known \emph{a
priori}. The CP decomposition of ${\boldsymbol{\cal Y}}$ can be
accomplished by solving:
\begin{equation}
\min _{\hat{\boldsymbol{A}}, \hat{\boldsymbol{B}},
\hat{\boldsymbol{C}}}\bigg\|{\boldsymbol{\cal Y}}-\sum_{u=1}^{U}
\hat{\boldsymbol{a}}_{u} \circ \hat{\boldsymbol{b}}_{u} \circ
\hat{\boldsymbol{c}}_{u}\bigg\|_{F}^{2}
\end{equation}
where we define $\hat{\boldsymbol{A}}\triangleq
[\hat{\boldsymbol{a}}_{1} \phantom{0} \ldots \phantom{0}
\hat{\boldsymbol{a}}_{U}]$, $\hat{\boldsymbol{B}}\triangleq
[\hat{\boldsymbol{b}}_{1} \phantom{0} \ldots \phantom{0}
\hat{\boldsymbol{b}}_{U}]$ and $\hat{\boldsymbol{C}}\triangleq
[\hat{\boldsymbol{c}}_{1} \phantom{0} \ldots \phantom{0}
\hat{\boldsymbol{c}}_{U}]$. The above optimization can be solved
by an ALS procedure which alternatively solves the following least
squares problems
\begin{align}
\hat{\boldsymbol{A}}^{(t+1)} &=\arg \min _{\hat{\boldsymbol{A}}}
\left\|\boldsymbol{Y}_{(1)}^{T}-\left(\hat{\boldsymbol{C}}^{(t)}
\odot \hat{\boldsymbol{B}}^{(t)}\right) \hat{\boldsymbol{A}}^{T}\right\|_{F}^{2} \\
\hat{\boldsymbol{B}}^{(t+1)} &=\arg \min _{\hat{\boldsymbol{B}}}
\left\|\boldsymbol{Y}_{(2)}^{T}-\left(\hat{\boldsymbol{C}}^{(t)}
\odot \hat{\boldsymbol{A}}^{(t+1)}\right) \hat{\boldsymbol{B}}^{T}\right\|_{F}^{2} \\
\hat{\boldsymbol{C}}^{(t+1)} &=\arg \min
_{\hat{\boldsymbol{C}}}\left\|\boldsymbol{Y}_{(3)}^{T}-\left(\hat{\boldsymbol{B}}^{(t+1)}
\odot \hat{\boldsymbol{A}}^{(t+1)}\right)
\hat{\boldsymbol{C}}^{T}\right\|_{F}^{2}
\end{align}
where $\boldsymbol{Y}_{(n)}$ denotes the mode-$n$ unfolding of the
tensor $\boldsymbol{\cal Y}$.

If the CP rank $U$ is unknown, a sparsity-promoting regularizer
can be added to the objective function such that one can
automatically determine the CP rank to find a low-rank
representation of the observed tensor. Let $\hat{U}>U$ denote an
overestimated CP rank. The problem can be formulated as
\cite{BazerqueMateos13,ZhouFang17}
\begin{align}
\mathop {\min
}\limits_{\boldsymbol{\hat{A}},\boldsymbol{\hat{B}},\boldsymbol{\hat{C}}}\quad
&\left\| {\boldsymbol{\mathcal Y} -\boldsymbol{\mathcal X}}
\right\|_F^2 + \mu
\left(\text{tr}({\boldsymbol{\hat{A}}}{{\boldsymbol{\hat{A}}}^H})
+ \text{tr}({\boldsymbol{\hat{B}}}{{\boldsymbol{\hat{B}}}^H}) +
\text{tr}({\boldsymbol{\hat{C}}}{{\boldsymbol{\hat{C}}}^H})\right) \nonumber\\
\text{s.t.}\quad &
\boldsymbol{\mathcal{X}}=\sum\limits_{l=1}^{\hat{U}}\boldsymbol{\hat{a}}_l\circ
\boldsymbol{\hat{b}}_l\circ\boldsymbol{\hat{c}}_l \label{opt-1}
\end{align}
where $\mu$ is a regularization parameter to control the tradeoff
between low-rankness and the data fitting error,
$\boldsymbol{\hat{A}}\triangleq
[\boldsymbol{\hat{a}}_1\phantom{0}\ldots\phantom{0}\boldsymbol{\hat{a}}_{\hat{U}}]$,
$\boldsymbol{\hat{B}}\triangleq
[\boldsymbol{\hat{b}}_1\phantom{0}\ldots\phantom{0}\boldsymbol{\hat{b}}_{\hat{U}}]$,
and $\boldsymbol{\hat{C}}\triangleq
[\boldsymbol{\hat{c}}_1\phantom{0}\ldots\phantom{0}\boldsymbol{\hat{c}}_{\hat{U}}]$.
Again, the above optimization problem can be solved via the ALS
method.


\subsection{CPD: A Vandermonde Structured-Based Approach}
The ALS-based CPD method involves a high computational complexity
as it needs to solve several large-scale least squares problems at
each iteration. Notice that, for our CPD problem, one of the
factor matrices is a Vandermonde matrix. Such a Vandermonde
structure along with its linear algebra properties can be utilized
to devise an efficient CPD method which yields a closed-form
solution of the factor matrices \cite{SorensenLathauwer13 ,LinJin19}. Details of the Vandermonde
structured-inspired CPD method are provided below.

To utilize the Vandermonde structure of the factor matrix
$\boldsymbol{C}$, we consider the mode-1 unfolding of the received
tensor ${\boldsymbol{\cal Y}}$, which is a matrix constructed by
concatenating the mode-1 fibers of ${\boldsymbol{\cal Y}}$ and can
be expressed as:
\begin{align}
\boldsymbol{Y}_{\left( 1 \right)}^T = \left( {\boldsymbol{C} \odot
\boldsymbol{B}} \right){\boldsymbol{A}^T}+ \boldsymbol{N}_{\left(
1 \right)}^T
\end{align}
We perform the truncated singular value decomposition (SVD)
$\boldsymbol{Y}_{\left( 1 \right)}^T =
\boldsymbol{U}{\mathbf{\Sigma}} {{\boldsymbol{V}}^H} \in\mathbb{C}
{^{T{N_{\rm s}}P \times Q}}$, where $\boldsymbol{U} \in
\mathbb{C}{^{T{N_{\rm s}}P \times U}}$, $\boldsymbol{\Sigma}\in
\mathbb{C}{^{ U \times  U}}$ and $\boldsymbol{V}\in \mathbb{C}{^{Q
\times U}}$.

To facilitate the exposition, we ignore the noise. Since
$(\boldsymbol{C} \odot \boldsymbol{B})$ is full column rank, we
know that there exists a nonsingular matrix $\boldsymbol{M} \in
\mathbb{C}{^{U \times U}}$ such that
\begin{align}\label{UM}
\boldsymbol{U M} = \boldsymbol{C} \odot \boldsymbol{B}
\end{align}
Consequently, we have
\begin{align} \label{U12}
\boldsymbol{U}_{1} \boldsymbol{M} &=  \underline{\boldsymbol{C}} \odot \boldsymbol{B} \\
\boldsymbol{U}_{2} \boldsymbol{M} &= \overline{\boldsymbol{C}}
\odot \boldsymbol{B}
\end{align}
where $\overline{\boldsymbol{A}}$ denotes a submatrix of
$\boldsymbol{A}$ obtained by removing the top row of
$\boldsymbol{A}$, $\underline{\boldsymbol{A}}$ denotes a submatrix
of $\boldsymbol{A}$ obtained by removing the bottom row of
$\boldsymbol{A}$, and
\begin{align}
\boldsymbol{U}_{1}=\boldsymbol{U}(1:(P-1) T{N_{\rm s}},:) \in \mathbb{C}^{(P-1) T{N_{\rm s}} \times U} \\
\boldsymbol{U}_{2}=\boldsymbol{U}(T{N_{\rm s}}+1: P T{N_{\rm
s}},:) \in \mathbb{C}^{(P-1) T{N_{\rm s}} \times U}
\end{align}
On the other hand, by utilizing the Vandermonde structure of
$\boldsymbol{C}$, we have
\begin{align} \label{CB}
\left( \underline{\boldsymbol{C}} \odot \boldsymbol{B}
\right)\boldsymbol{Z} = \overline{\boldsymbol{C}} \odot
\boldsymbol{B}
\end{align}
where $\boldsymbol{Z} \triangleq {\rm{diag}}\left(
{{z_1},\cdots,{z_ U}} \right)$, and ${{z_u} \triangleq {e^{ -
j2\pi {{{f_{\rm s}}}\over {{P_0}}}{\iota _u}}}}$ is the generator
of the factor matrix $\boldsymbol{C}$. Hence we arrive at
\begin{align}
{\boldsymbol{U}_2}\boldsymbol{M} =
{\boldsymbol{U}_1}\boldsymbol{MZ} \label{eqn1}
\end{align}
Since ${\underline{\boldsymbol{C}} \odot \boldsymbol{B}}$ is full
column rank, both ${\boldsymbol{U}_1}$ and ${\boldsymbol{U}_2}$
are full column rank. Therefore (\ref{eqn1}) can be further
rewritten as
\begin{align}
\boldsymbol{U}_1^{\dagger} {\boldsymbol{U}_2} = \boldsymbol{M}
\boldsymbol{Z}\boldsymbol{M}^{-1}
\end{align}
The above equation implies that the generators $\left\{ {{z_u}}
\right\}_{u = 1}^ U$ and $\boldsymbol{M}$ can be estimated from
the eigenvalue decomposition (EVD) of
$\boldsymbol{U}_1^{\dagger}\boldsymbol{U}_2$. Based on the
estimated generators $\{\hat{z}_u\}$, each column of the factor
matrix $\boldsymbol{C}$ can be estimated as
\begin{align}
{\boldsymbol{\hat c}_u} = \left[ {{{\hat z}_u}\phantom{0}{\hat
z}_u^2\phantom{0} \cdots\phantom{0} {\hat z}_u^P} \right]^T
\end{align}
According to (\ref{UM}), the column of the factor matrix
$\boldsymbol{B}$ can be estimated as
\begin{align}
{{\boldsymbol{\hat b}}_u} \triangleq \bigg( {{{{\boldsymbol{\hat
c}}_u^H} \over {{\boldsymbol{\hat c}}_u^H{{\boldsymbol{\hat
c}}_u}}} \otimes {{\boldsymbol{I}}_T}} \bigg)
{\boldsymbol{U}}{\boldsymbol{\hat M}}{{\left( {:,u} \right)}}
\end{align}
Finally, given $\boldsymbol{\hat B}$ and $\boldsymbol{\hat C}$,
the factor matrix $\boldsymbol{A}$ can be estimated as
\begin{align}
\boldsymbol{\hat A}=\boldsymbol{Y}_{(1)}\left((\boldsymbol{\hat C}
\odot \boldsymbol{\hat B})^{T}\right)^{\dagger}
\end{align}


\subsection{Channel Estimation}
After obtaining $\hat{\boldsymbol{A}}$, $\hat{\boldsymbol{B}}$ and
$\hat{\boldsymbol{C}}$, we now proceed to estimate the channel
parameters. From the CPD theory, we know that the estimated
$\{\hat{\boldsymbol{A}}
,\hat{\boldsymbol{B}},\hat{\boldsymbol{C}}\}$ and the true factor
matrices $\{{\boldsymbol{A}} ,{\boldsymbol{B}},{\boldsymbol{C}}\}$
are related as
\begin{align}
\hat{\boldsymbol{A}}
&=\boldsymbol{A}\boldsymbol{\Psi}_1\boldsymbol{\Gamma}+\boldsymbol{E}_1
\\
\hat{\boldsymbol{B}} &=\boldsymbol{B}\boldsymbol{ \Psi}_2 \boldsymbol{\Gamma}+\boldsymbol{E}_2 \\
\hat{\boldsymbol{C}} &=\boldsymbol{C}\boldsymbol{\Psi}_3
\boldsymbol{\Gamma}+\boldsymbol{E}_3
\end{align}
where $\{\boldsymbol{\Psi}_1, \boldsymbol{\Psi}_2,
\boldsymbol{\Psi}_3\}$ are nonsingular diagonal matrices which
satisfy $\boldsymbol{\Psi}_1
\boldsymbol{\Psi}_2\boldsymbol{\Psi}_3=\boldsymbol{I}_{U}$,
$\{\boldsymbol{E}_1, \boldsymbol{E}_2, \boldsymbol{E}_3\}$ are
estimation errors, and $\boldsymbol{\Gamma}$ is an unknown
permutation matrix. This permutation matrix $\boldsymbol{\Gamma}$
is common to all factor matrices, and thus can be ignored.

Recall that each column of the factor matrix $\boldsymbol{A}$ is
characterized by spatial angle parameters $\{{\zeta _u},{\xi
_u}\}$. Therefore these two spatial angle parameters can be
estimated through a correlation-based estimator:
\begin{align}
\{{\hat \zeta _u},{\hat \xi _u}\} = \arg \max_ {{{\zeta _u},{\xi
_u}}} \frac{{\left| {{\boldsymbol{\hat
a}}_u^H{{{\boldsymbol{\tilde a}}}_{{\rm{IRS}}}}\left( {{\zeta
_u},{\xi _u}} \right)} \right|}}{{{{\left\| {{{{\boldsymbol{\hat
a}}}_u}} \right\|}_2}{{\left\| {{{{\boldsymbol{\tilde
a}}}_{{\rm{IRS}}}}\left( {{\zeta _u},{\xi _u}} \right)}
\right\|}_2}}}
\end{align}
where ${{\boldsymbol{\hat a}}_u}$ denotes the $u$th column of
$\hat{\boldsymbol{A}}$.

Similarly, the spatial AoD associated with the BS ${\phi _u}$ and
the spatial AoA associated with the user ${\theta _u}$ can be
estimated as
\begin{align}
\{{\hat \phi_u},{\hat \theta _u}\} = \arg \max_{{\phi _u},{\theta
_u}} \frac{{\left| {{\boldsymbol{\hat b}}_u^H{{{\boldsymbol{\tilde
a}}}_{\rm{S}}}\left( {{\phi _u},{ \theta_u}} \right)}
\right|}}{{{{\| {{{{\boldsymbol{\hat b}}}_u}}\|}_2}{{\left\|
{{{{\boldsymbol{\tilde a}}}_{\rm{S}}}\left( {{\phi_u},{\theta _u}}
\right)} \right\|}_2}}}
\end{align}
where ${{\boldsymbol{\hat b}}_u}$ denotes the $u$th column of
$\hat{\boldsymbol{B}}$.

Note that the $u$th column of $\boldsymbol C$ is given by
$\boldsymbol{g}\left(\iota_u\right)$. Therefore the time delay
$\iota_u$ can be estimated via
\begin{align}
\hat{\iota}_{u}=\arg \min _{\iota_{u}}
\frac{\left|\hat{\boldsymbol{c}}_{u}^{H}
\boldsymbol{g}\left(\iota_{u}\right)\right|}{\left\|\hat{\boldsymbol{c}}_{u}\right\|_{2}
\left\|\boldsymbol{g}\left(\iota_{u}\right)\right\|_{2}}
\end{align}
where ${{\boldsymbol{\hat c}}_u}$ denotes the $u$th column of
$\hat{\boldsymbol{C}}$.


Next, we try to recover the composite path gains $\{
{\varrho}_u\}$. After obtaining $\{{\hat \zeta _u},{\hat \xi
_u}\}$, the factor matrix $\boldsymbol{A}$ can be accordingly
estimated as
\begin{align}
\boldsymbol{\tilde{A}}=\left[\tilde{\boldsymbol{a}}_{\rm{IRS}}({\hat
\zeta _1},{\hat \xi _1})\phantom{0} \cdots\phantom{0}
\tilde{\boldsymbol{a}}_{\rm{IRS}}({\hat \zeta _U},{\hat \xi
_U})\right]
\end{align}
Ignoring the estimation errors, $\boldsymbol{\hat{A}}$ and
$\boldsymbol{\tilde{A}}$ are related as
$\boldsymbol{\hat{A}}=\boldsymbol{\tilde{A}}\boldsymbol{\Psi}_1$.
Hence the nonsingular diagonal matrix $\boldsymbol{\Psi}_1$ can be
estimated as ${\boldsymbol \Psi}_1 = {\tilde {\boldsymbol
A}}^\dagger{\hat{ \boldsymbol A}}$. Similarly, after obtaining
$\{\hat{\iota}_{u}\}$, the factor matrix $\boldsymbol{C}$ can be
estimated as
\begin{align}
\boldsymbol{\tilde{C}}=\left[{\boldsymbol{g}}({\hat \iota
_1})\phantom{0} \cdots\phantom{0} {\boldsymbol{g}}({\hat \iota
_U})\right]
\end{align}
Thus, the nonsingular diagonal matrix $\boldsymbol{\Psi}_3$ can be
estimated as ${\boldsymbol \Psi}_3 = {\tilde {\boldsymbol
C}}^\dagger{\hat{ \boldsymbol C}}$. Since ${{\boldsymbol
\Psi}_1}{{\boldsymbol \Psi}_2}{{\boldsymbol \Psi}_3}={\mathbf
I}_U$, ${\boldsymbol \Psi}_2$ can be obtained as ${\boldsymbol
\Psi}_2 = \boldsymbol{\Psi}_{1}^{-1}\boldsymbol{\Psi}_{3}^{-1}$.


On the other hand, after obtaining $\{{\hat \phi_u},{\hat \theta
_u}\}$, we can construct a new matrix
\begin{align} \label{eqn7}
{\boldsymbol{\tilde B}} = \left[ {{{{\boldsymbol{\tilde
a}}}_{\rm{S}}}( {\hat \phi_1},{\hat \theta _1})\phantom{0}
\cdots\phantom{0}{{{\boldsymbol{\tilde a}}}_{\rm{S}}}( {\hat
\phi_U},{\hat \theta _U} )} \right]
\end{align}
Ideally we should have
$\boldsymbol{B}=\boldsymbol{\tilde{B}}\boldsymbol{D}$, where
$\boldsymbol{D}\triangleq\text{diag}(\varrho _{1},\cdots,\varrho
_{U})$. Moreover, ignoring estimation errors, we should have
$\boldsymbol{\hat{B}}=\boldsymbol{B}\boldsymbol{\Psi}_2$.
Therefore $\boldsymbol{D}$ can be estimated as
\begin{gather}
{\hat {\boldsymbol D}} = \boldsymbol{\tilde{B}}^{\dagger} {\hat
{\boldsymbol B}}\boldsymbol{\Psi}_2^{-1}
\end{gather}
Finally, the cascade channels $\{\boldsymbol{H}_p\}$ can be
estimated after those parameters $\{{\hat \zeta _u},{\hat \xi _u},
{\hat \phi_u},{\hat \theta _u}, {\hat \iota _u}, \hat {\varrho}_u
\}_{u=1}^U$ are obtained.

\section{Joint Active and Passive Beamforming Design} \label{sec:beamforming}
In this section, we consider the problem of joint active and
passive beamforming design based on the estimated channel
parameters. Specifically, we aim to maximize the spectral
efficiency by jointly optimizing the precoding matrices at the
transmitter, the combining matrices at the receiver, and the
reflection coefficients at the RIS. In the data transmission
stage, the received signal associated with the $p$th subcarrier
can be expressed as
\begin{align}
\boldsymbol{\bar y}_{p}&=\boldsymbol{W}_{\mathrm{BB}, p}^{H}
\boldsymbol{W}_{\mathrm{RF}}^{H} {{\boldsymbol{R}}_p}{{\boldsymbol{\Phi }}}{{\boldsymbol{G}}_p}
\boldsymbol{F}_{\mathrm{RF}} \boldsymbol{F}_{\mathrm{BB}, p} {\boldsymbol{s}_p}
+\boldsymbol{W}_{\mathrm{BB}, p}^{H}
\boldsymbol{W}_{\mathrm{RF}}^{H}\boldsymbol{n}_{p}\nonumber\\
&= \boldsymbol{W}_{\mathrm{BB}, p}^{H}
\boldsymbol{W}_{\mathrm{RF}}^{H} {{\boldsymbol{\bar H}}_p}
\boldsymbol{F}_{\mathrm{RF}} \boldsymbol{F}_{\mathrm{BB}, p}
{\boldsymbol{s}_p}+\boldsymbol{\bar n}_{p}
\end{align}
where ${{\boldsymbol{\bar H}}_p}\triangleq
{{\boldsymbol{R}}_p}{{\boldsymbol{\Phi
}}}{{\boldsymbol{G}}_p}\in\mathbb{C}^{N_{\rm r}\times N_{\rm t}}$
is the equivalent (i.e. effective) channel associated with the
$p$th subcarrier, and
$\boldsymbol{n}_{p}\sim\mathcal{CN}(0,\sigma^2\boldsymbol{I})$
denotes the additive white Gaussian noise.

Assuming that the transmitted signal obeys a Gaussian
distribution, the achievable spectral efficiency can be calculated
as
\begin{align}
R&=\frac{1}{P}\sum_{p=1}^{P} \log _{2}
\operatorname{det}(\boldsymbol{I}_{N_{\rm s}}+
\frac{1}{\sigma^{2}}\left(\boldsymbol{W}_{\mathrm{RF}}\boldsymbol{W}_{\mathrm{BB},p}
\right)^{\dagger}
\boldsymbol{\bar H}_{p} \nonumber\\
& \ \ \ \ \ \times \boldsymbol{F}_{\mathrm{RF}}
\boldsymbol{F}_{\mathrm{BB}, p} \boldsymbol{F}_{\mathrm{BB},
p}^{H} \boldsymbol{F}_{\mathrm{RF}}^{H} \boldsymbol{\bar
H}_{p}^{H}\left(\boldsymbol{W}_{\mathrm{RF}}\boldsymbol{W}_{\mathrm{BB},p}
\right))
\end{align}
where $\dagger$ denotes the Moore-Penrose inverse.

\subsection{Problem Formulation}
With the objective of maximizing the spectral efficiency, the
joint beamforming problem can be formulated as
\begin{align} \label{original optimization problem}
\max_{\cal S} & \frac{1}{P}\sum_{p=1}^{P} \log _{2}
\operatorname{det} \left(\boldsymbol{I}_{N_{\rm
s}}+\frac{1}{\sigma^{2}}
\left(\boldsymbol{W}_{\mathrm{RF}}\boldsymbol{W}_{\mathrm{BB},p}
\right)^{\dagger}
\boldsymbol{\bar H}_{p} \boldsymbol{F}_{\mathrm{RF}} \right.\nonumber\\
&\ \ \ \ \ \ \ \ \ \ \ \ \ \ \ \left.\times
\boldsymbol{F}_{\mathrm{BB}, p}\boldsymbol{F}_{\mathrm{BB}, p}^{H}
\boldsymbol{F}_{\mathrm{RF}}^{H} \boldsymbol{\bar H}_{p}^{H}
\left(\boldsymbol{W}_{\mathrm{RF}}\boldsymbol{W}_{\mathrm{BB},p}\right)\right) \nonumber\\
\text { s.t. } &\left\|\boldsymbol{F}_{\mathrm{RF}}
\boldsymbol{F}_{\mathrm{BB}, p}\right\|_{F}^{2}
\leq \rho, \forall p=1, \ldots, P \nonumber\\
&\left|\boldsymbol{F}_{\mathrm{RF}}(i, j)\right|=
\left|\boldsymbol{W}_{\mathrm{RF}}(i, j)\right|=1, \forall i, j \nonumber\\
& \boldsymbol{\bar H}_{p}=\boldsymbol{R}_{p} \boldsymbol{\Phi}
\boldsymbol{G}_{p}, \forall p=1, \ldots, P \nonumber\\
& \mathbf{\Phi}=\operatorname{diag}(e^{j \varsigma_{1}}, e^{j
\varsigma_{2}}, \cdots, e^{j \varsigma_{M}})
\end{align}
where ${\cal
S}\triangleq\left\{\boldsymbol{F}_{\mathrm{RF}},\left\{\boldsymbol{F}_{\mathrm{BB},
p}\right\}_{p=1}^{P},
\boldsymbol{W}_{\mathrm{RF}}\left\{\boldsymbol{W}_{\mathrm{BB},
p}\right\}_{p=1}^{P}, \boldsymbol{\Phi}\right\}$ is the set of
optimization variables, and $\rho$ is the transmit power
constraint. To simplify the problem, we first ignore the
constraint introduced by the hybrid analog/digital structure and
consider a fully digital precoder/combiner. Let $\boldsymbol{F}_p
\in \mathbb{C}^{N_{\rm t} \times N_{\rm s}}$ and $\boldsymbol{W}_p
\in \mathbb{C}^{N_{\rm r} \times N_{\rm s}}$ denote a fully
digital precoder and a fully digital combiner, respectively. The
problem (\ref{original optimization problem}) can be simplified as
\begin{align}\label{C1 optimization problem}
\max _{\left\{\left\{\boldsymbol{F}_{p}\right\}_{p=1}^{P},\left\{\boldsymbol{W}_{p}\right\}_{p=1}^{P},
\boldsymbol{\Phi}\right\}} & \frac{1}{P}\sum_{p=1}^{P} \log _{2} \operatorname{det}
\left(\boldsymbol{I}_{N_{\rm s}}+\frac{1}{\sigma^{2}}
\boldsymbol{W}_{p}^{\dagger} \boldsymbol{\bar H}_{p}\right.\nonumber\\
&\ \ \ \ \ \ \ \ \ \ \ \ \left.\times\boldsymbol{F}_{p} \boldsymbol{F}_{p}^{H}
\boldsymbol{\bar H}_{p}^{H} \boldsymbol{W}_{p}\right)\nonumber \\
\text { s.t. } &\left\|\boldsymbol{F}_{p}\right\|_{F}^{2} \leq \rho, \forall p=1, \ldots, P\nonumber \\
& \boldsymbol{\bar H}_{p}=\boldsymbol{R}_{p} \boldsymbol{\Phi}
\boldsymbol{G}_{p}, \forall p=1, \ldots, P \nonumber\\
& \mathbf{\Phi}=\operatorname{diag}(e^{j \varsigma_{1}}, e^{j
\varsigma_{2}}, \cdots, e^{j \varsigma_{M}})
\end{align}
Once an optimal fully digital precoder/combiner is found, we can
use the manifold optimization method \cite{kasaiHiroyuki18} to
search for a hybrid precoder/combiner to approximate the optimal
fully digital precoder/combiner. Due to the sparse scattering
nature of mmWave channels, such a strategy has been proven
effective in previous studies, e.g. \cite{ElRajagopal14, YuShen16}, which showed that hybrid
beamforming/combining with a small number of RF chains can
asymptotically approach the performance of fully digital
beamforming/combining.



Given the reflection matrix $\boldsymbol{\Phi}$, we first study
how to devise the fully digital precoder/combiner. Let
$r_p={\mathrm{rank}}\left({\boldsymbol{\bar H}_{p}}\right)$.
Define the truncated SVD of the effective channel
$\boldsymbol{\bar H}_{p}$ as
\begin{align}
\boldsymbol{\bar H}_{p}&=\boldsymbol{U}_{p} \boldsymbol{\Sigma}_{p} \boldsymbol{V}_{p}^{H}\nonumber\\
&=\left[
\begin{array}{ll}
\boldsymbol{U}_{1,p} & \boldsymbol{U}_{2,p}
\end{array}
\right] \left[
\begin{array}{cc}
\boldsymbol{\Sigma}_{1,p} & \mathbf{0}\\
\mathbf{0} & \boldsymbol{\Sigma}_{2,p}
\end{array}
\right] \left[
\begin{array}{ll}
\boldsymbol{V}_{1,p} & \boldsymbol{V}_{2,p}
\end{array}
\right]^{H}
\end{align}
where $\boldsymbol{U}_{p}\in\mathbb{C}^{N_{\rm r} \times r_p}$,
$\boldsymbol{V}_{p}\in\mathbb{C}^{N_{\rm t}\times r_p}$, and
$\boldsymbol{\Sigma}_p$ is an $r_p \times r_p$ diagonal matrix.
Also, we have $\boldsymbol{U}_{1,p}\in\mathbb{C}^{N_{\rm r} \times
N_{\rm s}}$, $\boldsymbol{\Sigma}_{1,p}\in\mathbb{C}^{N_{\rm s}
\times N_{\rm s}}$ and $\boldsymbol{V}_{1,p}\in\mathbb{C}^{N_{\rm
t} \times N_{\rm s}}$. Given a specified $\boldsymbol{\Phi}$, the
optimal fully digital precoder/combiner is given as
\begin{align}
\boldsymbol{F}_{\mathrm{opt},p}=\boldsymbol{V}_{1,p}
\boldsymbol{\Lambda}_{p}^{1 / 2}, \quad
\boldsymbol{W}_{\mathrm{opt},p}=\boldsymbol{U}_{1,p}
\end{align}
where
\begin{align}
\boldsymbol{\Lambda}_{p} = \text{diag}\left(\rho_{1,p}, . . . ,
\rho_{N_{\rm s},p} \right)
\end{align}
and
\begin{align}
\rho_{i,p}=\max (1 / \lambda_p-{{\sigma^{2}}/
\left({\boldsymbol{\Sigma}}_{1,p}(i, i)\right)^{2}}, 0), i=1,
\ldots, N_{{\rm s}}
\end{align}
denotes the optimal amount of power allocated to the $i$th data
stream, $1/{\lambda_p}$ is the water level satisfying
$\sum_{i=1}^{N_{\rm s}}\rho_{i,p}=\rho$. Thanks to the massive
array gain provided by the RIS, the effective signal-to-noise
ratio (SNR) is usually large, in which case an equal power
allocation scheme is near-optimal. Therefore we can approximate
$\boldsymbol{F}_{\mathrm{opt},p}$ as:
\begin{align}
\boldsymbol{F}_{\text{opt},p} \approx \sqrt{\rho / N_{{\rm s}}}
\boldsymbol{V}_{1,p}
\end{align}

Substituting the optimal precoder/combiner
$\{\boldsymbol{F}_{\text{opt},p}\}$ and
$\{\boldsymbol{W}_{\text{opt},p}\}$ into (\ref{C1 optimization
problem}), we arrive at a problem which concerns only the
optimization of the passive beamforming matrix
$\boldsymbol{\Phi}$:
\begin{align} \label{C2 optimization problem}
\max _{\boldsymbol{\Phi}} &\quad \frac{1}{P}\sum_{p=1}^{P} \log
_{2} \operatorname{det}\left(\boldsymbol{I}_{N_{\rm
s}}+\frac{\rho}{N_{{\rm s}} \sigma^{2}}
\left(\boldsymbol{\Sigma}_{1,p}\right)^{2}\right) \nonumber\\
\text { s.t. } &\quad \mathbf{\Phi}=\operatorname{diag}(e^{j
\varsigma_{1}}, e^{j \varsigma_{2}}, \cdots, e^{j \varsigma_{M}}).
\end{align}

\subsection{Passive Beamforming Design}
We assume $\left|\alpha_{1}\right| \geq\left|\alpha_{2}\right|
\geq \ldots \geq\left|\alpha_{L}\right|$, $\left|\beta_{1}\right|
\geq\left|\beta_{2}\right| \geq \ldots \geq\left|\beta_{L_{\rm
r}}\right|$. To gain insight into the passive beamforming design,
we write the equivalent channel as follows:
\begin{align}
\boldsymbol{\bar H}_{p}=& \boldsymbol{R}_{p} \boldsymbol{\Phi} \boldsymbol{G}_{p} \nonumber\\
=&\left(\sum_{n=1}^{L_{\rm r}} \beta_{n} e^{-j 2 \pi f_{{\rm s}} \kappa_{n}
\frac{p}{P_{0}}} \boldsymbol{a}_{\rm{UE}}\left(\theta_{n}\right) \boldsymbol{a}_{\rm{IRS}}^{H}
\left(\vartheta_{n}^{{\rm t}}, \chi_{n}^{{\rm t}}\right)\right)\boldsymbol{\Phi} \nonumber\\
& \times\left(\sum_{m=1}^{L} \alpha_{m} e^{-j 2 \pi f_{{\rm s}}
\tau_{m} \frac{p}{P_{0}}}
\boldsymbol{a}_{\rm{IRS}}\left(\vartheta_{m}^{{\rm r}},
\chi_{m}^{{\rm r}}\right)
\boldsymbol{a}^H_{\rm{BS}}\left(\phi_{m}\right) \right) \nonumber\\
=& \sum_{m=1}^{L} \sum_{n=1}^{L_{\mathrm{{\rm r}}}} \alpha_{m} \beta_{n} e^{-j 2 \pi f_{s}
\left(\tau_{m}+\kappa_{n}\right) \frac{p}{P_{0}}} \boldsymbol{a}_{\mathrm{UE}}\left(\theta_{n}\right)\nonumber\\
&\ \ \ \ \ \ \ \ \ \ \ \ \ \times
\underbrace{{\boldsymbol{a}_{\rm{IRS}}^{H}\left(\vartheta_{n}^{{\rm
t}}, \chi_{n}^{{\rm
t}}\right)\boldsymbol{\Phi}\boldsymbol{a}_{\rm{IRS}}\left(\vartheta_{m}^{{\rm
r}},
\chi_{m}^{{\rm r}}\right)}}_{d_{m n}} \boldsymbol{a}_{\rm{BS}}^H\left(\phi_{m}\right) \nonumber\\
=& \sum_{m=1}^{L} \sum_{n=1}^{L_{{\rm r}}} \alpha_{m} \beta_{n} e^{-j 2 \pi f_{s}
\left(\tau_{m}+\kappa_{n}\right){\frac{p}{P_{0}}}} d_{m n} \boldsymbol{a}_{\rm{UE}}
\left(\theta_{n}\right) \boldsymbol{a}_{\mathrm{BS}}^{H}\left(\phi_{m}\right) \nonumber\\
=& \boldsymbol{A}_{\mathrm{UE}} \boldsymbol{D}_{p}
\boldsymbol{A}^H_{\mathrm{BS}} \label{eqn2}
\end{align}
where we define $\boldsymbol{D}_{p}\left(m,n\right)\triangleq
\alpha_{m} \beta_{n} e^{-j 2 \pi f_{{\rm
s}}\left(\tau_{m}+\kappa_{n}\right){\frac{p}{P_{0}}}} d_{m n} $,
and
\begin{align}
d_{m n}
&\triangleq{\boldsymbol{a}_{\rm{IRS}}^{H}\left(\vartheta_{n}^{{\rm
t}}, \chi_{n}^{{\rm
t}}\right)\boldsymbol{\Phi}\boldsymbol{a}_{\rm{IRS}}\left(\vartheta_{m}^{{\rm
r}},
\chi_{m}^{{\rm r}}\right)} \nonumber\\
&=\boldsymbol{v}^{H}\left(\boldsymbol{a}^{*}_{\rm{IRS}}\left(\vartheta_{n}^{{\rm
t}}, \chi_{n}^{{\rm t}}\right) \circ
\boldsymbol{a}_{\rm{IRS}}\left(\vartheta_{m}^{{\rm r}},
\chi_{m}^{{\rm r}}\right)\right) \nonumber\\
&=\boldsymbol{v}^{H}
\boldsymbol{a}_{\mathrm{IRS}}\left(\vartheta_{m}^{{\rm
r}}-\vartheta_{n}^{{\rm t}}, \chi_{m}^{{\rm r}}-\chi_{n}^{{\rm t}}
\right)
\end{align}
\begin{align}
\boldsymbol{A}_{\mathrm{UE}} &\triangleq \left[\boldsymbol{a}_{\mathrm{UE}}
\left(\theta_{1}\right)\phantom{0}
\cdots\phantom{0} \boldsymbol{a}_{\mathrm{UE}}\left(\theta_{L_{\rm r}}\right)\right] \\
\boldsymbol{A}_{\mathrm{BS}} &\triangleq
\left[\boldsymbol{a}_{\mathrm{BS}}\left(\phi_{1}\right)\phantom{0}
\cdots\phantom{0}
\boldsymbol{a}_{\mathrm{BS}}\left(\phi_{L}\right)\right]
\end{align}
In the above equation, $\alpha_m\beta_n$ is the complex gain
associated with the $(m,n)$th BS-RIS-user composite path which is
composed of the $m$th path from the BS to the RIS and the $n$th
path from the RIS to the user, $\{\vartheta_{m}^{{\rm
r}}-\vartheta_{n}^{{\rm t}},\chi_{m}^{{\rm r}}-\chi_{n}^{{\rm
t}}\}$ are the composite RIS angles associated with the $(m,n)$th
composite path, and $d_{mn}$ is referred to as the passive
beamforming gain associated with the $(m,n)$th composite path.
Clearly, there are a total number of $U=LL_r$ composite paths.

When the numbers of antennas $N_{\rm t}$ and $N_{\rm r}$ are
sufficiently large, $\boldsymbol{A}_{\mathrm{UE}}$ and
$\boldsymbol{A}_{\mathrm{BS}}$ can be considered as orthonormal
matrices with unit-norm and mutually orthogonal columns. If the
phase shift vector $\boldsymbol{v}$ is properly devised such that
the off-diagonal elements of $\boldsymbol{D}_p$ are small relative
to entries on the main diagonal, then $\boldsymbol{\bar H}_{p} =
\boldsymbol{A}_{\mathrm{UE}} \boldsymbol{D}_{p}
\boldsymbol{A}^H_{\mathrm{BS}}$ can be approximated as a truncated
SVD of $\boldsymbol{\bar H}_{p}$, in which case the optimization
problem (\ref{C2 optimization problem}) turns into
\begin{align}\label{C5 optimization problem}
\max_{\boldsymbol{v}} &\quad \frac{1}{P}\sum_{p=1}^{P}
\sum_{i=1}^{N_{\rm s}} \log _{2}
\left(1+\frac{\rho}{N_{\rm s} \sigma^{2}}\left|\boldsymbol{D}_{p}(i, i)\right|^{2}\right) \nonumber\\
\text { s.t. } &\quad \boldsymbol{D}_{p}(i, i)=\alpha_{i}
\beta_{i} e^{-j 2 \pi f_{\rm s}
\left(\tau_{i}+\kappa_{i}\right)} d_{i i}  \nonumber\\
&\quad d_{i i}=\boldsymbol{v}^{H} \boldsymbol{a}_{\mathrm{IRS}}
\left(\vartheta_{i}^{\rm r}-\vartheta_{i}^{\rm t}, \chi_{i}^{\rm r}-\chi_{i}^{\rm t}\right)  \nonumber\\
&\quad \left|d_{i j}\right|=|\boldsymbol{v}^{H}
\boldsymbol{a}_{\mathrm{IRS}}\left(\vartheta_{i}^{\rm
r}-\vartheta_{j}^{\rm t}, \chi_{i}^{\rm r}-\chi_{j}^{\rm t}
\right)|
<\delta, \forall i \neq j  \nonumber\\
&\quad \boldsymbol{v}=\left[e^{j \varsigma_{1}}\phantom{0} e^{j
\varsigma_{2}}\phantom{0} \cdots\phantom{0} e^{j
\varsigma_{M}}\right]^{H}
\end{align}
where $\delta$ is a small positive value and the constraint
$|d_{ij}|<\delta$ is imposed to ensure that $\boldsymbol{\bar
H}_{p} = \boldsymbol{A}_{\mathrm{UE}} \boldsymbol{D}_{p}
\boldsymbol{A}^H_{\mathrm{BS}}$ is a good approximation of the
truncated SVD of $\boldsymbol{\bar H}_{p}$. As analyzed in
\cite{WangFang21}, due to the asymptotic orthogonality of RIS's
array response vectors characterized with different angular
parameters, the constraint $|d_{ij}|<\delta$ can be neglected and
the solution to the simplified problem can automatically guarantee
that off-diagonal entries of $\boldsymbol{D}_p$ are small relative
to entries on its main diagonal. Specifically, by ignoring the
constraint $|d_{ij}|<\delta$, the optimization (\ref{C5
optimization problem}) can be simplified as
\begin{align}\label{C3 optimization problem}
\max_{\boldsymbol{v}} &\quad \frac{1}{P}\sum_{p=1}^{P}
\sum_{i=1}^{N_{\rm s}} \log _{2}\left(1+\frac{\rho}{N_{\rm s}
\sigma^{2}}\left|\alpha_{i} \beta_{i}\right|^{2}
\boldsymbol{v}^{H} \boldsymbol{P}_{ii} \boldsymbol{v}\right) \nonumber\\
\text { s.t. } & \quad \boldsymbol{v}=\left[e^{j
\varsigma_{1}}\phantom{0} e^{j \varsigma_{2}}\phantom{0}
\cdots\phantom{0} e^{j \varsigma_{M}}\right]^{H}
\end{align}
where $\boldsymbol{P}_{ii}\triangleq
\boldsymbol{a}_{\mathrm{IRS}}\left(\vartheta_{i}^{\rm
r}-\vartheta_{i}^{\rm t}, \chi_{i}^{\rm r}-\chi_{i}^{\rm t}
\right)\boldsymbol{a}^{H}_{\mathrm{IRS}}\left(\vartheta_{i}^{\rm
r}-\vartheta_{i}^{\rm t}, \chi_{i}^{\rm r}-\chi_{i}^{\rm t}
\right)$. Such an optimization can be efficiently solved via a
manifold optimization technique, whose details can be found in
\cite{WangFang21}.

From (\ref{C3 optimization problem}), we see that to optimize the
reflection coefficients, we only need the knowledge of the
composite gains $\{\alpha_i\beta_i\}_{i=1}^{N_s}$ and the
composite RIS angles $\{\vartheta_{i}^{\rm r}-\vartheta_{i}^{\rm
t}, \chi_{i}^{\rm r}-\chi_{i}^{\rm t}\}_{i=1}^{N_s}$. Recall that,
in the channel estimation stage, the following channel parameters
$\{{\hat \zeta _u},{\hat \xi _u}, {\hat \phi_u},{\hat \theta _u},
{\hat \iota _u}, \hat {\varrho}_u \}_{u=1}^U$ are obtained, in
which we have
\begin{gather}
\nonumber
{{\alpha _m}{\beta _n} \mapsto {\varrho _u}, \quad u = 1, \ldots ,L{L_{\rm r}}}\\
\vartheta _m^{\rm r} - \vartheta _n^{\rm t} \mapsto \zeta _u,
\quad u = 1, \ldots ,L{L_{\rm r}}\\\nonumber \chi _m^{\rm r} -
\chi _n^{\rm t} \mapsto \xi _u, \quad u = 1, \ldots ,L{L_{\rm r}}
\end{gather}
We see that our proposed channel estimator can provide an estimate
of the composite gains as well as the composite RIS angles
associated with all $U$ composite paths. The problem now is how to
appropriately choose $N_s$ composite paths from these $U$
composite paths. Randomly choosing $N_s$ composite paths certainly
does not work. In fact, from (\ref{eqn2}), it is easy to know that
the composite paths corresponding to the diagonal entries of
$\boldsymbol{D}_p$ must have mutually distinct AoDs at the BS and
mutually distinct AoAs at the user. Also, to improve the spectral
efficiency, clearly we should choose those composite paths whose
composite gains are as large as possible. Based on the above
considerations, the $N_s$ composite paths can be selected based on
the following criterion:
\begin{align}
\max _{\mathcal{I}} &\quad \sum_{i \in \mathcal{I}} |\hat{\varrho}_{i}|^{2}  \nonumber\\
\text { s.t. } &\quad \mathcal{I} \subset\{1, \cdots, U\}, \quad|\mathcal{I}|=N_{\rm s},  \nonumber\\
&\quad \left|\boldsymbol{a}_{\mathrm{BS}}^{H}(\hat \phi_{i})
\boldsymbol{a}_{\mathrm{BS}}
(\hat \phi_{j})\right|<\delta_{\mathrm{BS}}, i \neq j, \forall i, j \in \mathcal{I}  \nonumber\\
& \quad \left|\boldsymbol{a}_{\mathrm{UE}}^{H}(\hat \theta_{i})
\boldsymbol{a}_{\mathrm{UE}}(\hat
\theta_{j})\right|<\delta_{\mathrm{UE}}, i \neq j, \forall i, j
\in \mathcal{I},
\end{align}
where the last two constraints are imposed to ensure that the
selected composite paths have mutually distinct AoDs at the BS and
mutually distinct AoAs at the user, in which
$\delta_{\mathrm{BS}}$ and $\delta_{\mathrm{UE}}$ are small
positive parameters of user's choice. Based on the selected $N_s$
composite paths, the optimization problem (\ref{C3 optimization
problem}) can be further written as
\begin{align}
\max _{\boldsymbol{v}} &\quad \sum_{i \in \mathcal{I}} \log _{2}
\left(1+\frac{\rho \hat{\varrho}_{i}^{2}}{N_{\rm s} \sigma^{2}}
\boldsymbol{v}^{H}
\boldsymbol{P}_{ii} \boldsymbol{v}\right)  \nonumber\\
\text { s.t. } &\quad
\boldsymbol{P}_{ii}=\boldsymbol{a}_{\operatorname{IRS}} ({\hat
\zeta _i},{ \hat \xi _i}) \boldsymbol{a}_{\mathrm{IRS}}^{H}
({\hat \zeta _i},{ \hat \xi _i}), \forall i \in \mathcal{I}  \nonumber\\
& \quad \boldsymbol{v}=\left[e^{j \varsigma_{1}}, e^{j
\varsigma_{2}}, \cdots, e^{j \varsigma_{M}}\right]^{H}
\end{align}
The above optimization problem can be efficiently solved via the
manifold optimization method proposed in \cite{kasaiHiroyuki18}.

\subsection{Active Beamforming Design}
After the passive beamforming vector $\boldsymbol{v}$ is
determined, according to (\ref{eqn2}), the $p$th subcarrier's
equivalent channel $\boldsymbol{\bar{H}}_p$ can be estimated as
\begin{align}
{\boldsymbol{\hat {\bar{H}}}}_{p} &=\sum_{u=1}^{U}
\hat{\varrho}_{u} {e^{-j 2 \pi f_{{\rm
s}}{\hat{\iota}}_{u}\frac{p}{P_{0}}}} \boldsymbol{v}^H
\boldsymbol{a}_{\mathrm{IRS}}({\hat \zeta _u},{ \hat \xi _u})\nonumber\\
&\ \ \ \ \ \ \ \ \ \ \ \ \ \ \ \ \ \ \ \ \ \ \ \ \ \ \ \times\boldsymbol{a}_{\mathrm{UE}}
(\hat \theta_{u}) \boldsymbol{a}_{\mathrm{BS}}^{H}(\hat \phi_{u}) \nonumber\\
&=\left[\begin{array}{ll} \hat{\boldsymbol U}_{1, p} &
\hat{\boldsymbol{U}}_{2, p}
\end{array}\right]\left[\begin{array}{cc}
\hat{\boldsymbol{\Sigma}}_{1, p} & \mathbf{0} \\
\mathbf{0} & \hat{\mathbf{\Sigma}}_{2, p}
\end{array}\right]\left[\begin{array}{ll}
\hat{\boldsymbol{V}}_{1, p} & \hat{\boldsymbol{V}}_{2, p}
\end{array}\right]^{H}
\end{align}
where the second equality is a truncated SVD of ${\boldsymbol{\hat
{\bar{H}}}}_{p}$. Based on the previous discussion, the optimal
fully digital precoder/combiner can be obtained as
\begin{equation}
\boldsymbol{F}_{\mathrm{opt}, p}^{*}=\sqrt{\rho / N_{\rm s}}
\boldsymbol{\hat V}_{1,p}, \quad \boldsymbol{W}_{\mathrm{opt},
p}^{*}=\boldsymbol{\hat U}_{1,p}
\end{equation}

After the optimal fully digital precoder/combiner is obtained, we
search for a common analog precoding (combining) matrix
$\boldsymbol{F}_{\text{RF}}$ ($\boldsymbol{W}_{\text{RF}}$) and a
set of baseband precoding (combining) matrices
$\{\boldsymbol{F}_{\text{BB},p}\}$
($\{\boldsymbol{W}_{\text{BB},p}\}$) to approximate the optimal
precoder (combiner) $\{\boldsymbol{F}_{\text{opt},p}\}$
($\{\boldsymbol{W}_{\text{opt},p}\}$). The problem can be
formulated as
\begin{align}\label{C4 optimization problem1}
\min_{\boldsymbol{F}_{\mathrm{RF}},\left\{\boldsymbol{F}_{\mathrm{BB},
p}\right\}_{p=1}^{p}} &\quad
\sum_{p=1}^{P}\left\|\boldsymbol{F}_{\mathrm{opt},
p}^{\star}-\boldsymbol{F}_{\mathrm{RF}}
\boldsymbol{F}_{\mathrm{BB}, p}\right\|_{F}^{2} \quad \nonumber\\
\text { s.t. } &\quad \left|\boldsymbol{F}_{\mathrm{RF}}(i,
j)\right|=1, \forall i, j,
\end{align}
\begin{align} \label{C4 optimization problem2}
\min
_{\boldsymbol{W}_{\mathrm{RF}},\left\{\boldsymbol{W}_{\mathrm{BB},
p}\right\}_{p=1}^{p}} & \quad
\sum_{p=1}^{P}\left\|\boldsymbol{W}_{\mathrm{opt},
p}^{\star}-\boldsymbol{W}_{\mathrm{RF}}
\boldsymbol{W}_{\mathrm{BB}, p}\right\|_{F}^{2} \quad \nonumber\\
\text { s.t. } &\quad \left|\boldsymbol{W}_{\mathrm{RF}}(i,
j)\right|=1, \forall i, j,
\end{align}
The above optimization problem can be solved via the manifold
optimization technique introduced in \cite{AbsilMahony09}.

%
%

\section{Simulation Results} \label{sec:results}
We present simulation results to evaluate the performance of the
proposed CPD-based channel estimation methods and the joint
beamforming scheme. In Section \ref{sec:CPD-method}, two different
approaches are introduced to perform the CPD, namely, the ALS
method and the Vandemonde structure-based method. The
corresponding channel estimation methods are referred to as
CPD-ALS and CPD-VS, respectively.

In our simulations, we assume that the BS employs a ULA with
$N_t=32$ antennas and $R_t=4$ RF chains, the IRS is equipped with
$M=16 \times 16$ passive reflecting elements, and the user employs
a ULA with $N_r=32$ and $R_r=4$ RF chains. The angular parameters
$\left\{\vartheta_{l}^{{\rm r}}, \chi_{l}^{{\rm
r}},\phi_{l}\right\}_{l=1}^{L}$, $\left\{\vartheta_{l}^{{\rm t}},
\chi_{l}^{{\rm t}},\theta_{l}\right\}_{l=1}^{L_r}$ are randomly
generated from $[0,2\pi]$, where we set $L=3$ and $L_r=3$. The
delay spreads
$\left\{\tau_l\right\}_{l=1}^{L},\left\{\kappa_l\right\}_{l=1}^{L_r}$
are drawn from a uniform distribution ${\cal{U}}(0,100{\rm{ns}})$.
The complex gains $\left\{\alpha_l\right\}_{l=1}^{L}$ and
$\left\{\beta_l\right\}_{l=1}^{L_r}$ follow a circularly symmetric
Gaussian distribution ${\cal{CN}}(0,1)$. The number of data
streams is set to $N_s=2$. The total number of subcarriers is set
to $P_0=128$, among which $P$ subcarriers are used for training.
The sampling rate is set to $f_s=0.32\rm GHz$. The signal-to-noise
ratio (SNR) is defined as
\begin{align}
\text{SNR} \triangleq \frac{\|\boldsymbol{\cal { Y
}}-\boldsymbol{\cal{N}}\|_{F}^{2}}{\|\boldsymbol{\cal{N
}}\|_{F}^{2}}
\end{align}

\begin{figure*}[ht]
\centering
\includegraphics[scale=0.37]{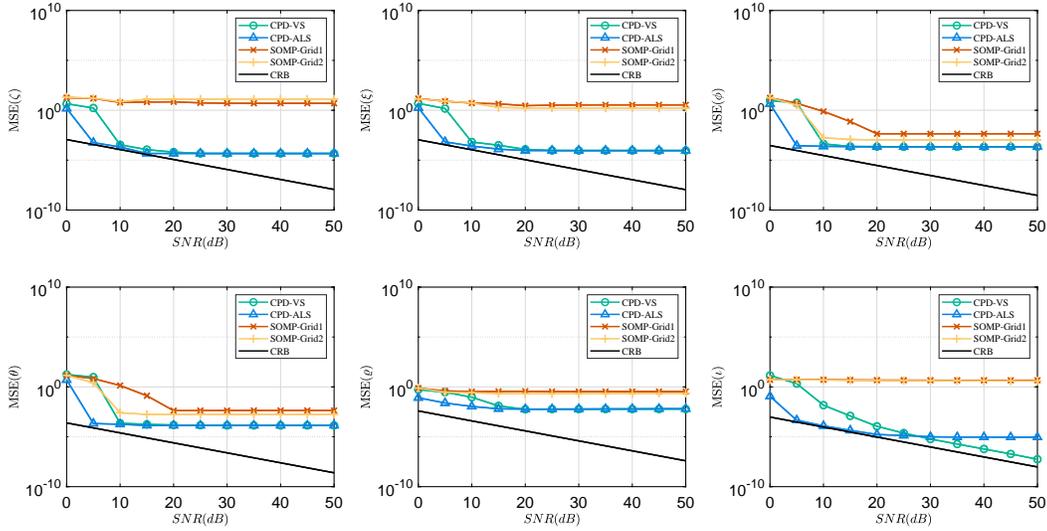}
\caption{MSEs of respective methods vs. SNR, where $Q=16$, $T=16$,
and $P=16$.} \label{fig3}
\end{figure*}

\begin{figure*}[ht]
 \centering
 \subfigure[NMSEs vs SNR]{\label{fig:subfig:a}
\includegraphics[scale=0.39]{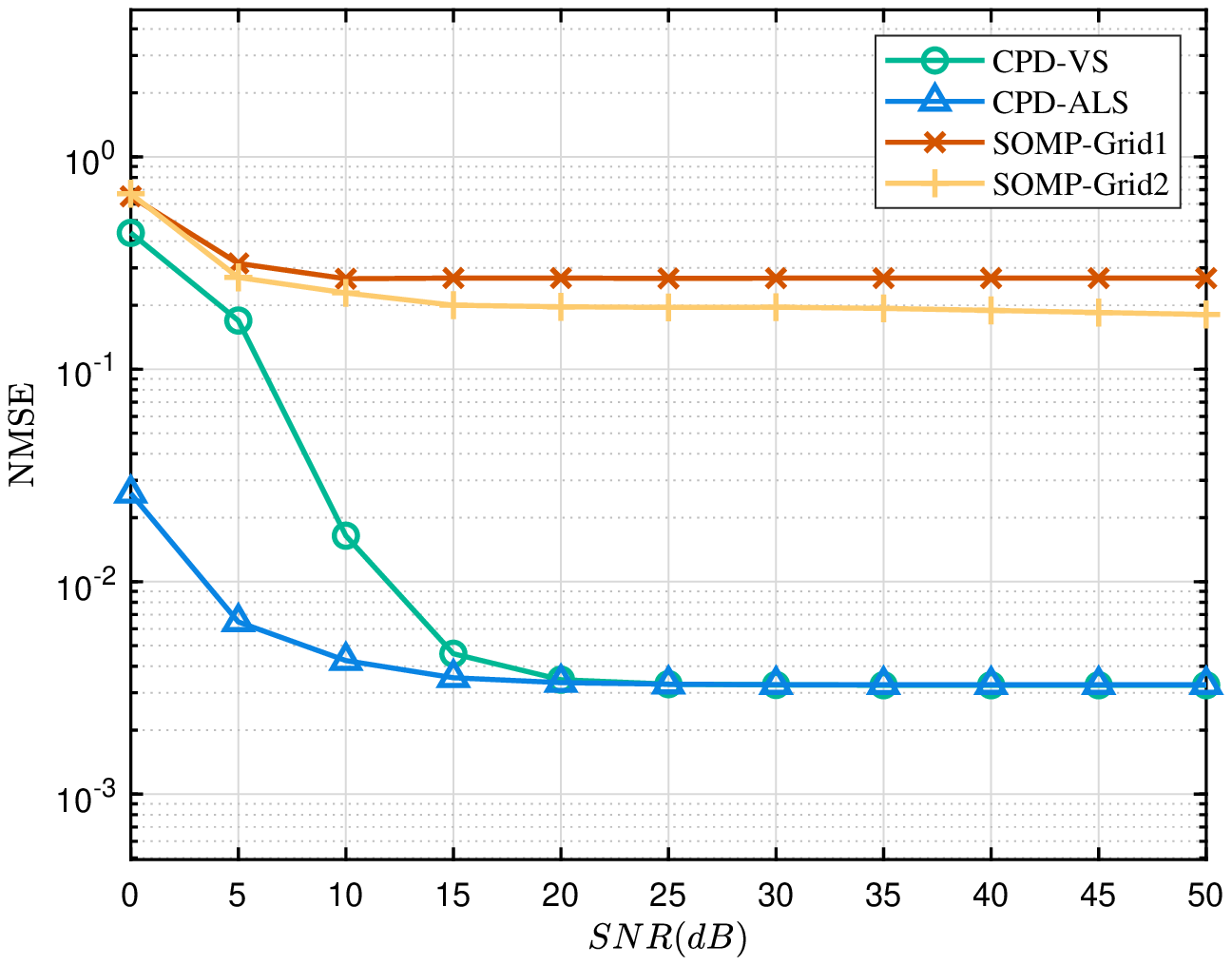}}
\hspace*{-5ex}
\subfigure[NMSEs vs $P$]{\label{fig:subfig:b}
\includegraphics[scale=0.39]{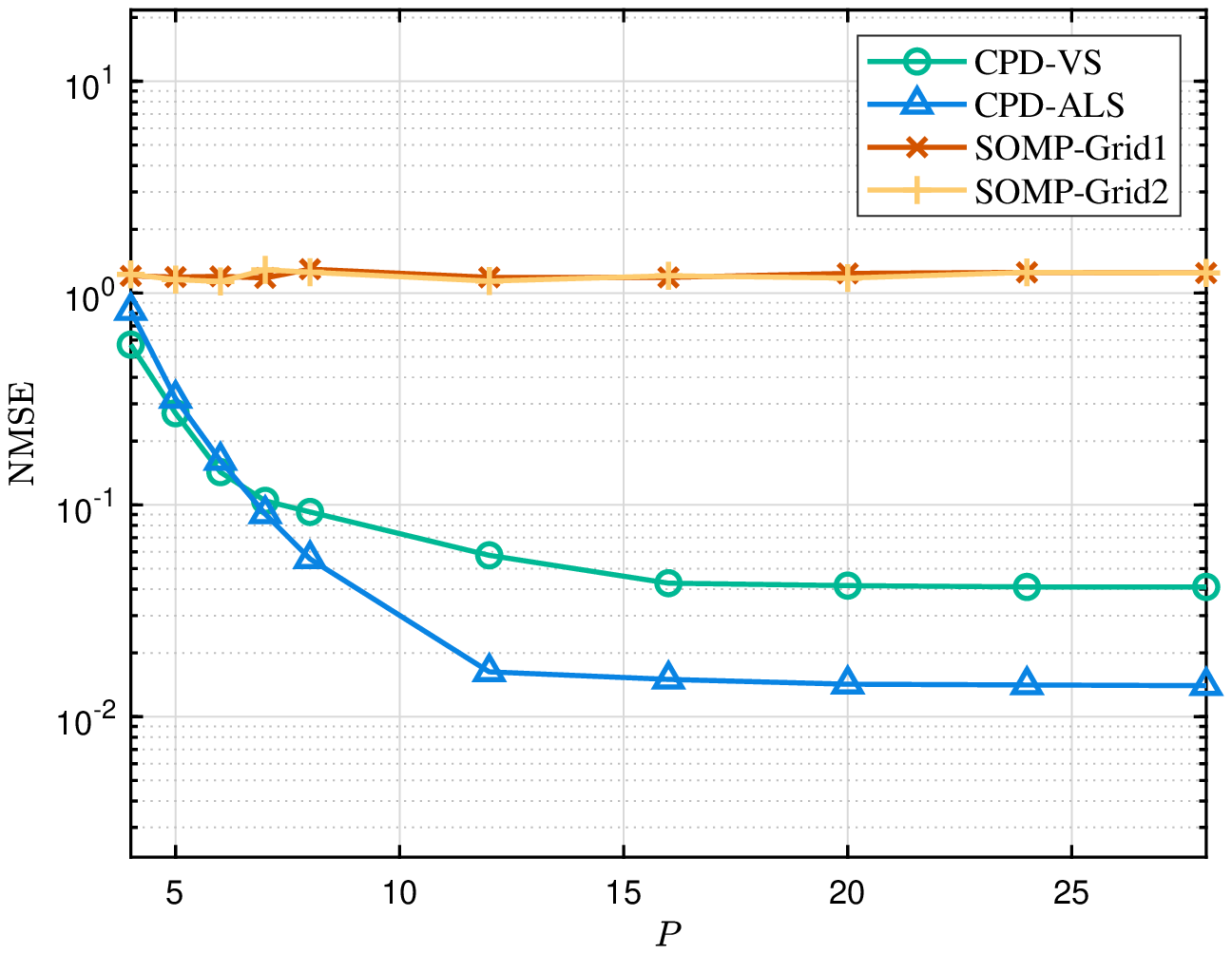}}
\hspace*{-5ex}
\subfigure[NMSEs vs $T$]{\label{fig:subfig:c}
\includegraphics[scale=0.39]{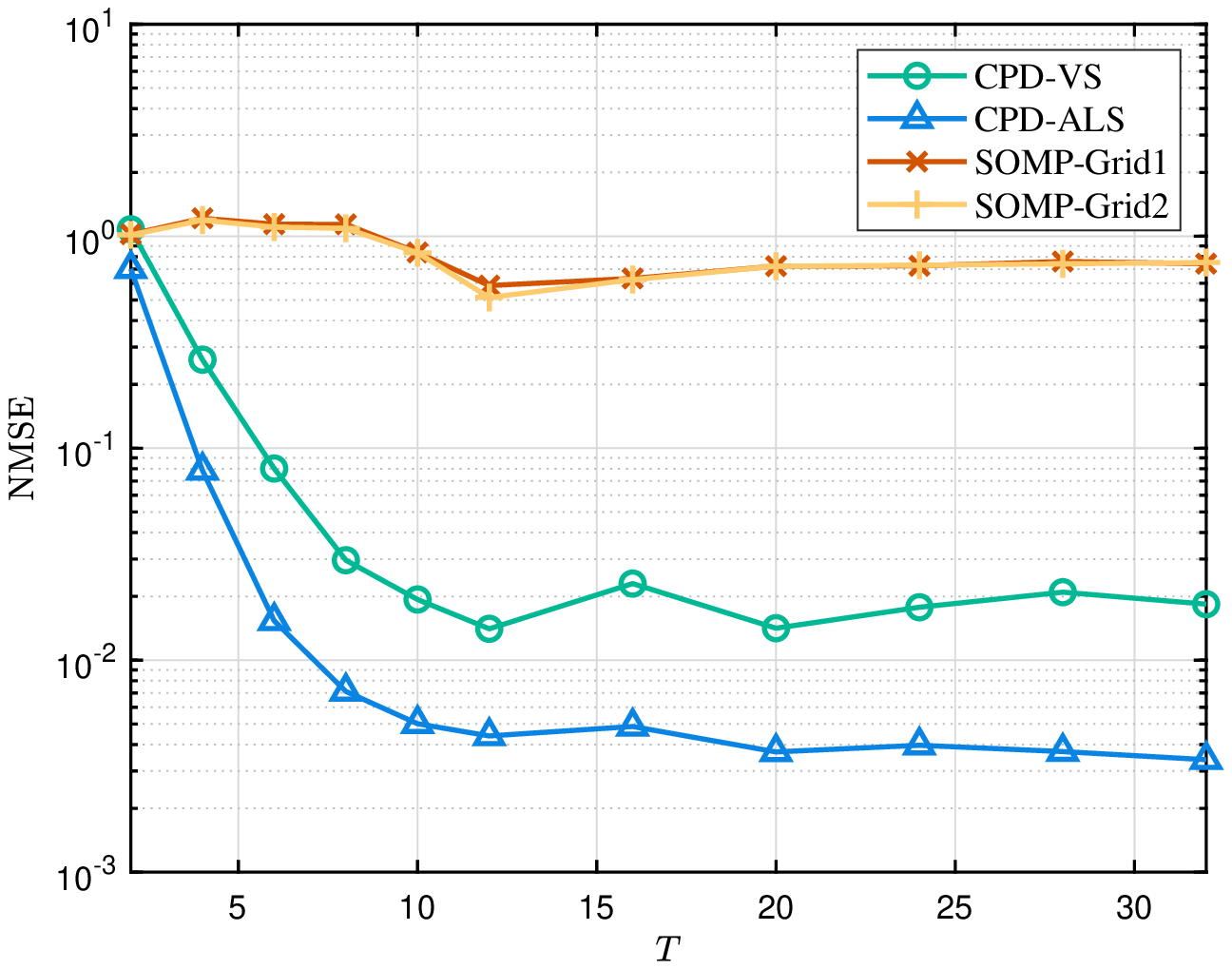}}
\caption{(a) NMSEs of respective methods vs. SNR, where $Q=16$,
$T=16$, and $P=16$; (b) NMSEs of respective methods vs. $P$, where
$Q=10$, $T=5$, and $\text{SNR}=20$dB; (c) NMSEs of respective
methods vs. $T$, where $Q=10$, $P=10$, and $\text{SNR}=20$dB.}
\label{fig4}
\end{figure*}

\begin{figure*}[ht]
 \centering
\subfigure[Spectral efficiency vs SNR]{\label{fig:subfig:a}
\includegraphics[scale=0.57]{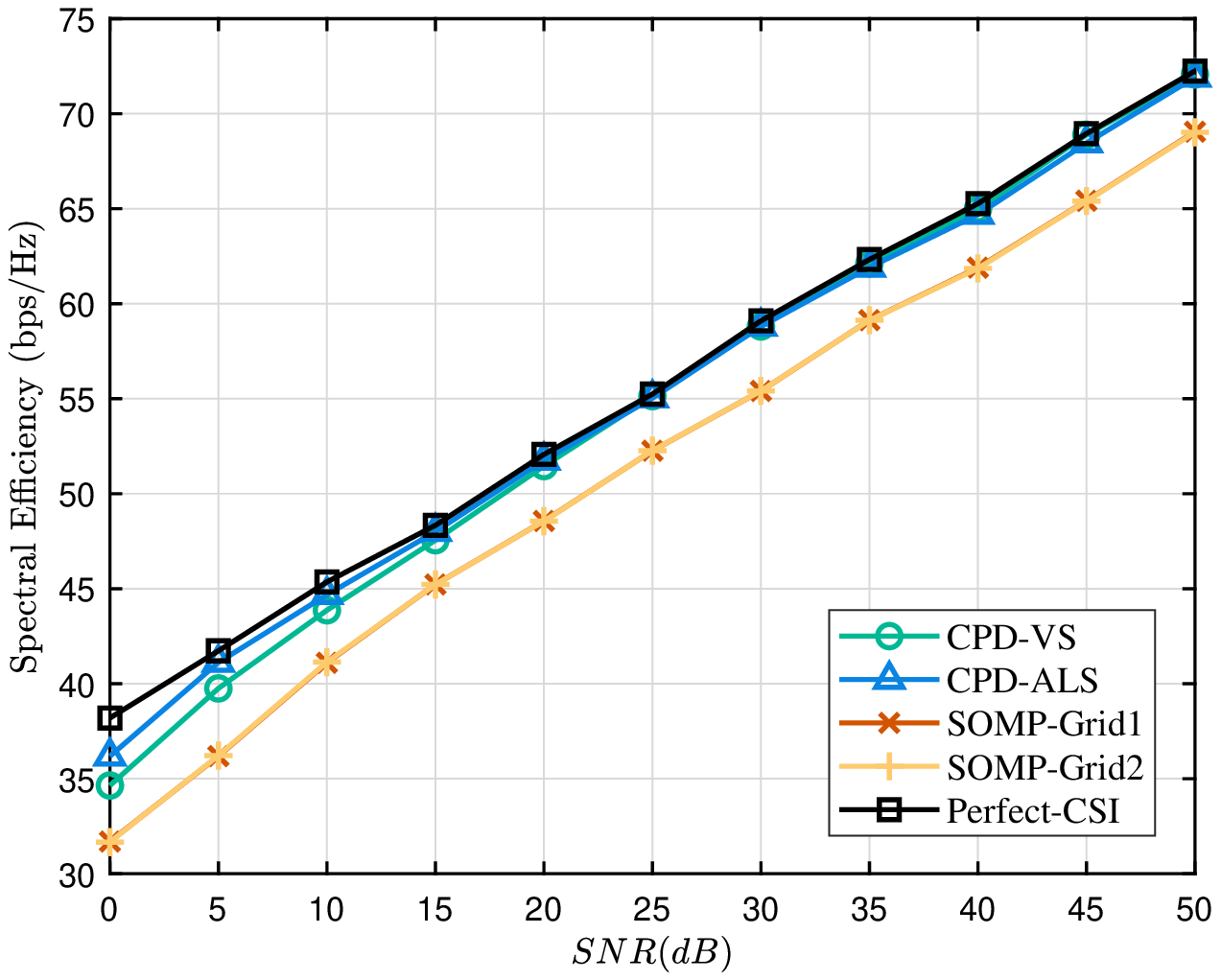}}
\hspace*{-6ex} \subfigure[Spectral efficiency vs
$T$]{\label{fig:subfig:b}
\includegraphics[scale=0.57]{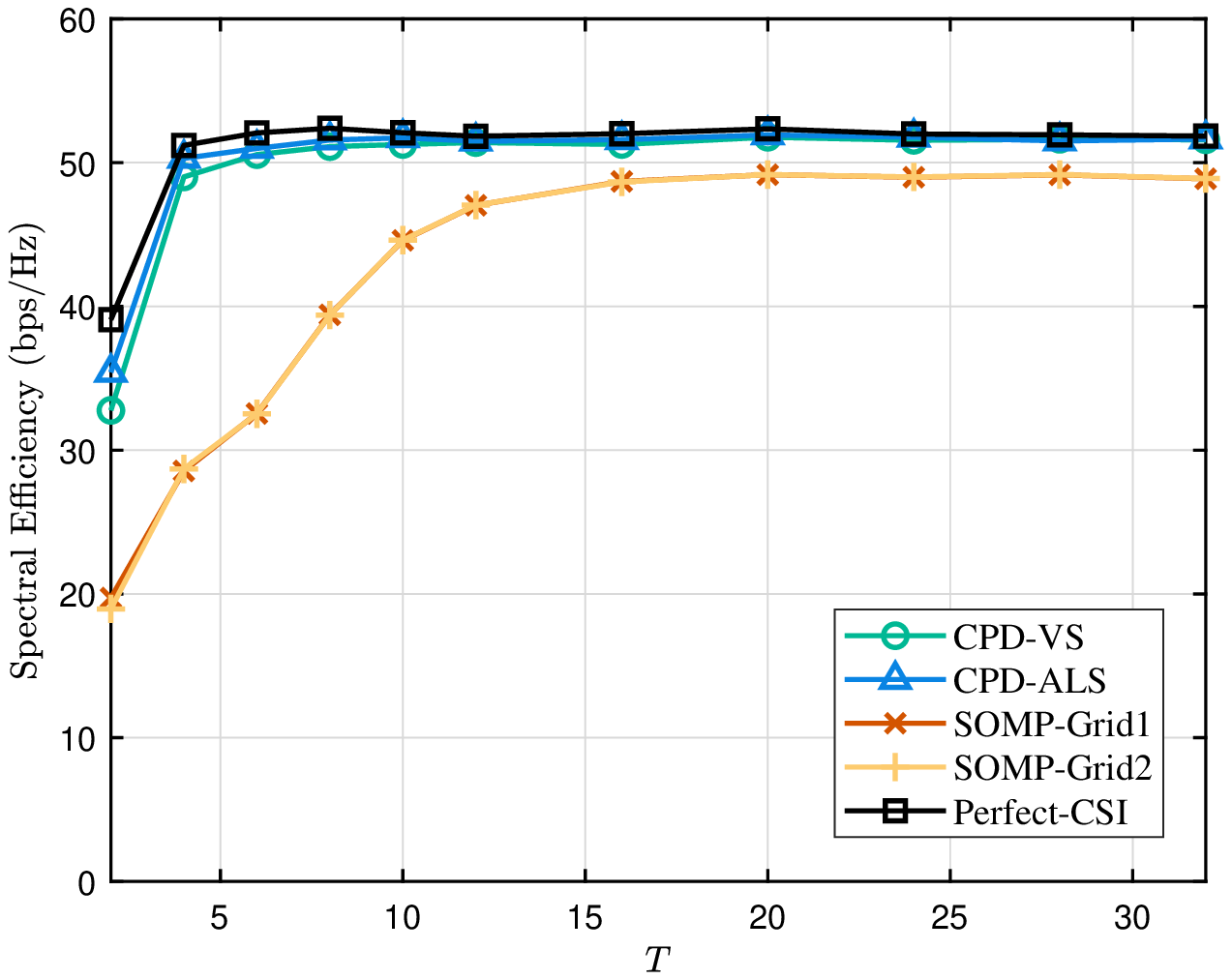}}
\caption{(a) Spectral efficiency of respective methods vs. SNR,
where $Q=16$, $T=16$, and $P=16$; (b) Spectral efficiency of
respective methods vs. $T$, where $Q=16$, $P=16$, and
$\text{SNR}=20$dB.} \label{fig5}
\end{figure*}

We first examine the estimation accuracy of the channel parameters
$\left\{ \zeta _u, \xi _u, \phi _u, \theta _u, \iota _u, \varrho
_u\right\}_{u = 1}^{ U}$ and the overall channel estimation
performance. Note that the channel estimation problem being
considered in this work can be cast as a multi-measurement vector
(MMV) compressed sensing problem, and the simultaneous-OMP method
(SOMP) \cite{TroppGilbert06} can be used to estimate the cascade
channel. For the SOMP method, two different grids are employed to
discretize the continuous parameter space: the first grid
discretizes the multi-dimensional parameter space into $128 \times
\left(128 \times 128 \right) \times 128 \times 128$ points, and
the second grid discretizes the continuous parameter space into
$256 \times \left(256 \times 256 \right) \times 256\times 256$
points. The CRB results are also included to provide a benchmark
for evaluating the performance of our proposed method. The
calculation of CRB can be found in Appendix \ref{appA}.




In Fig. \ref{fig3}, we plot the mean square errors (MSEs) of the
estimated channel parameters as a function of the SNR, where we
set $P=16$, $T=16$, and $Q=16$. We see that our proposed methods
can achieve an estimation accuracy close to the theoretical lower
bound. Also, the CPD-ALS method is superior to the CPD-VS method
in the low SNR regime. This is probably because the CPD-VS which
depends critically on the structure of the factor matrix is more
sensitive to noise. Also, it can be observed that both CPD-based
methods present a substantial advantage over the SOMP method.

In Fig. \ref{fig4}(a), we plot the estimation performance of
respective methods as a function of the SNR. The performance is
evaluated via the normalized mean squared error (NMSE) of the
cascaded channel, which is defined as
$\sum_{p=1}^{P}\|\boldsymbol{\hat
H}_{p}-\boldsymbol{H}_{p}\|_F^2/\sum_{p=1}^{P}\|\boldsymbol{H}_{p}\|_F^2$.
Again, we see that our proposed methods present a significant
performance improvement over the SOMP method. In particular, the
CPD-ALS method achieves a decent estimation performance even in a
low SNR regime, say $\text{SNR}=0$dB. Note that in mmWave
communications, due to the severe path loss, the SNR for channel
estimation is usually low, with the range of interest from 0dB to
10dB at most. Thus the ability of delivering an accurate channel
estimate in the low SNR regime is highly desirable and has
important practical implications.


In Fig. \ref{fig4}(b), we plot the estimation performance of
respective methods as a function of the number of subcarriers $P$,
where we set $Q=10$ and $T=5$. It can be seen that the proposed
methods provide a reliable channel estimate when $P\geq 8$, which
corresponds to a total number of $400$ measurements for training.
As a comparison, note that the cascade channel $\boldsymbol{H}_p$
to be estimated has a size of $N_t N_r\times M=1024\times 256$,
which has more than $2.6\times 10^5$ parameters. This result
indicates that the proposed methods can achieve a substantial
training overhead reduction. Fig. \ref{fig4}(c) plots the
estimation performance versus the number of time slots $T$, where
we set $Q=10$ and $P=10$. This result, again, demonstrates the
superiority of the proposed methods over the compressed
sensing-based method.


Next, we examine the beamforming performance attained by the joint
beamforming scheme proposed in Section \ref{sec:beamforming}. To
illustrate the effectiveness of the proposed channel estimator, we
include the beamforming performance attained by assuming the
perfect knowledge of the CSI, which serves as an upper bound on
the beamforming performance attained by using the estimated CSI.
Fig. \ref{fig5} plots the spectral efficiency of the proposed
joint beamforming scheme as a function of SNR and the number of
time frames, respectively. We see that our proposed CPD-based
estimators achieve performance close to that attained by assuming
perfect CSI knowledge even in the low SNR regime, which verifies
the effectiveness of the proposed estimation method. Also, our
proposed methods present a clear performance improvement over the
compressed sensing-based method. Particularly, when the training
overhead is low, say, $P=Q=16$ and $T=2$, the proposed CPD-based
methods can still achieve decent beamforming performance, whereas
the compressed sensing-based method incurs a significant
performance loss.

\section{Conclusions}
In this paper, by exploiting the intrinsic multi-dimensional
structure as well as the sparse scattering characteristics of the
mmWave channels, we developed two CPD-aided channel estimation
methods, namely, an ALS-based CPD method and a Vandemonde
structure-based CPD method, for RIS-assisted mmWave MIMO-OFDM
systems. The proposed methods effectively utilize the low-rankness
of the CPD formulation and can achieve a substantial training
overhead reduction. We also developed a joint beamforming scheme
that utilizes the estimated cascade channel parameters for
optimizing the system's active and passive variables. Simulation
results show that our proposed methods present a significant
performance advantage over the compressed sensing method, and can
achieve superior channel estimation and beamforming performance
with a low training overhead.

\section{Derivation of Cram\'{e}r-Rao Lower Bound} \label{appA}
Consider the $Q \times TN_{\rm s} \times P$ observation tensor
$\boldsymbol{{\cal Y}}$ in (\ref{eqn18})
\begin{align}
\boldsymbol{{\cal Y}} = \sum\limits_{u = 1}^U
{{{{\boldsymbol{\tilde a}}}_{{\rm{IRS}}}}} \left( {{\zeta
}_{u}},{{\xi }_{u}} \right) \circ \left( {{\varrho
_u}{\boldsymbol{\tilde a}}_{\rm{S}}\left( {{\phi _u},{\theta _u}}
\right)} \right) \circ {\boldsymbol{g}}\left( {{\iota _u}} \right)
+ \boldsymbol{{\cal N}}
\end{align}
where $\boldsymbol{{\cal N}}\left( {q,t,p} \right) \sim {\cal
C}{\cal N}\left( {0,{\sigma ^2}} \right)$, $\left\{ {{\zeta
}_{u}},{{\xi }_{u}},{\phi _u},{\theta _u},{\varrho _u}, {\iota
_u}\right\}$ are the unknown channel parameters to be estimated.
Let $\boldsymbol{p}\triangleq[\boldsymbol{\zeta }^T,
\boldsymbol{\xi }^T,\boldsymbol{\phi}^T, \boldsymbol{\theta}^T,
\boldsymbol{\varrho }^T,\boldsymbol{\iota }^T]$, where
\[\begin{array}{*{20}{c}}
{\begin{array}{*{20}{c}} {{{\boldsymbol{\zeta }}} \triangleq
{{\left[ {\begin{array}{*{20}{c}} {{\zeta _{1}}}\phantom{0} \cdots
\phantom{0} {{\zeta _{U}}}
\end{array}} \right]}^T}}&
{{{\boldsymbol{\xi }}} \triangleq{{\left[ {\begin{array}{*{20}{c}}
{{\xi_{1}}}\phantom{0} \cdots \phantom{0} {{\xi _{U}}}
\end{array}} \right]}^T}}
\end{array}}\\
{\begin{array}{*{20}{c}} {{\boldsymbol{\phi }} \triangleq {{\left[
{\begin{array}{*{20}{c}} {{\phi_1}}\phantom{0} \cdots \phantom{0}
{{\phi_U}}
\end{array}} \right]}^T}}&{\boldsymbol{\theta}  \triangleq {{\left[ {\begin{array}{*{20}{c}}
{{\theta_1}}\phantom{0} \cdots \phantom{0} {{\theta _U}}
\end{array}} \right]}^T}}
\end{array}}\\
{\begin{array}{*{20}{c}} {{\boldsymbol{\varrho }} \triangleq
{{\left[ {\begin{array}{*{20}{c}} {{\varrho _1}}\phantom{0} \cdots
\phantom{0} {{\varrho _U}}
\end{array}} \right]}^T}}
&{{\boldsymbol{\iota }} \triangleq {{\left[
{\begin{array}{*{20}{c}} {{\iota _1}}\phantom{0} \cdots
\phantom{0} {{\iota _U}}
\end{array}} \right]}^T}}
\end{array}}
\end{array}\]
Thus the log-likelihood function of $\boldsymbol p$ can be
expressed as
\begin{align}
L\left( {\boldsymbol{p}} \right) &= f\left( {{\boldsymbol {\cal Y}};{\boldsymbol{A}},
{\boldsymbol{B}},{\boldsymbol{C}}} \right)\nonumber\\
 &=  - QT{N_s}P\ln \left( {\pi {\sigma ^2}} \right) - \frac{1}{{{\sigma ^2}}}\left\|
 {{\boldsymbol{Y}}_{\left( 1 \right)}^T - \left( {{\boldsymbol{C}} \odot {\boldsymbol{B}}}
 \right){{\boldsymbol{A}}^T}} \right\|_F^2\nonumber\\
 &=  - QT{N_s}P\ln \left( {\pi {\sigma ^2}} \right) - \frac{1}{{{\sigma ^2}}}\left\|
 {{\boldsymbol{Y}}_{\left( 2 \right)}^T - \left( {{\boldsymbol{C}} \odot {\boldsymbol{A}}}
 \right){{\boldsymbol{B}}^T}} \right\|_F^2\nonumber\\
 &=  - QT{N_s}P\ln \left( {\pi {\sigma ^2}} \right) - \frac{1}{{{\sigma ^2}}}\left\|
 {{\boldsymbol{Y}}_{\left( 3 \right)}^T - \left( {{\boldsymbol{B}} \odot {\boldsymbol{A}}}
 \right){{\boldsymbol{C}}^T}} \right\|_F^2
\end{align}
The complex Fisher information matrix (FIM) for $\boldsymbol p$ is
given by
\begin{align}
\Omega ({\boldsymbol{p}}) = \mathbb{E}\left\{ {{{\left(
{\frac{{\partial L({\boldsymbol{p}})}}{{\partial
{\boldsymbol{p}}}}} \right)}^H}\left( {\frac{{\partial
L({\boldsymbol{p}})}}{{\partial {\boldsymbol{p}}}}} \right)}
\right\}
\end{align}
To calculate $\Omega \left({\boldsymbol{p}}\right)$, we first
compute the partial derivative of $L\left({\boldsymbol{p}}\right)$
with respect to ${\boldsymbol{p}}$ and then calculate the
expectation with respect to $p\left(\boldsymbol{{\cal Y}};
\boldsymbol p\right)$.

\subsection{Partial Derivative of $L(\boldsymbol{p})$  W.R.T $\boldsymbol{p}$}
For simplicity, we consider the partial derivative of
$L(\boldsymbol{p})$ with respect to ${{\zeta }_{u}}$. Partial
derivations of $L(\boldsymbol{p})$ with respect to other
parameters can be deduced in a similar way and thus omitted. We
have
\begin{align}
\frac{\partial L(\boldsymbol{p})}{\partial {{\zeta
}_{u}}}=\text{tr}\left\{ {{\left( \frac{\partial
L(\boldsymbol{p})}{\partial \boldsymbol{A}}
\right)}^{T}}\frac{\partial \boldsymbol{A}}{\partial {{\zeta
}_{u}}}+{{\left( \frac{\partial L(\boldsymbol{p})}{\partial
{{\boldsymbol{A}}^{*}}} \right)}^{T}}\frac{\partial
{{\boldsymbol{A}}^{*}}}{\partial {{\zeta }_{u}}} \right\}
\end{align}
where
\begin{align}
  \frac{\partial L(\boldsymbol{p})}{\partial \boldsymbol{A}}&=
  \frac{1}{{{\sigma }^{2}}}{{\left( \boldsymbol{Y}_{\left( 1 \right)}^{T}-
  \left(\boldsymbol{C}\odot\boldsymbol{B} \right){{\boldsymbol{A}}^{T}}
  \right)}^{H}}\left( \boldsymbol{C}\odot \boldsymbol{B} \right)
\\
  \frac{\partial L(\boldsymbol{p})}{\partial {{\boldsymbol{A}}^{*}}}
  &=\frac{1}{{{\sigma }^{2}}}{{\left( \boldsymbol{Y}_{\left( 1 \right)}^{T}-
  \left( \boldsymbol{C}\odot \boldsymbol{B} \right){{\boldsymbol{A}}^{T}}
  \right)}^{T}}{{\left( \boldsymbol{C}\odot \boldsymbol{B} \right)}^{*}}
\\
\frac{\partial \boldsymbol{A}}{\partial {{\zeta }_{u}}}&=\left[
  \boldsymbol 0 \phantom{0}
\cdots \phantom{0}   {{{\boldsymbol{\tilde{a}}}}_{a,u}}\phantom{0}
\cdots \phantom{0}  \boldsymbol 0 \right]
\end{align}
in which ${{\boldsymbol{\tilde{a}}}_{a,u}}\triangleq
j{{\boldsymbol{V}}^{T}}{{\boldsymbol{D}}_{1}}{{\boldsymbol{a}}_{\text{IRS}}}\left(
{{\zeta}_{u}},{{\xi}_{u}} \right)$ and
\begin{align}
\boldsymbol{D}_{1} \triangleq \operatorname{diag}(\underbrace{0,
\cdots, 0}_{M_{\rm{z}}}, \underbrace{1, \cdots, 1}_{M_{\rm{z}}},
\cdots, \underbrace{M_{\mathrm{y}}-1, \cdots,
M_{\mathrm{y}}-1}_{M_{\rm{z}}})
\end{align}
Thus we have
\begin{small}
\begin{align}
  \frac{\partial L(\boldsymbol{p})}{\partial {{\zeta }_{u}}}&=\boldsymbol{e}_{u}^{T}
  \frac{1}{{{\sigma }^{2}}}{{\left( \boldsymbol{C}\odot\boldsymbol{B} \right)}^{T}}{{
  \left( \boldsymbol{Y}_{\left( 1 \right)}^{T}-\left(\boldsymbol{C}\odot \boldsymbol{B}
  \right){{\boldsymbol{A}}^{T}} \right)}^{*}}{{{\boldsymbol{\tilde{a}}}}_{a,u}}\nonumber\\
  &\ \ +\boldsymbol{e}_{u}^{T}\frac{1}{{{\sigma }^{2}}}{{\left( \boldsymbol{C}\odot\boldsymbol{B}
  \right)}^{H}}\left(\boldsymbol{Y}_{\left( 1 \right)}^{T}-\left(\boldsymbol{C}\odot\boldsymbol{B}
  \right){{\boldsymbol{A}}^{T}} \right)\boldsymbol{\tilde{a}}_{a,u}^{*} \nonumber\\
 & =2\operatorname{Re}\left\{ \boldsymbol{e}_{u}^{T}\frac{1}{{{\sigma }^{2}}}{{\left(
 \boldsymbol{C}\odot \boldsymbol{B} \right)}^{T}}{{\left( \boldsymbol{Y}_{\left( 1
\right)}^{T}-\left( \boldsymbol{C}\odot \boldsymbol{B}
\right){{\boldsymbol{A}}^{T}}
\right)}^{*}}{{{\boldsymbol{\tilde{A}}}}_{a}}{{\boldsymbol{e}}_{u}}
\right\}
\end{align}
\end{small}
where $\operatorname{Re}\left\{ \cdot  \right\}$ represents the
real part of a complex number, ${{\boldsymbol{e}}_{u}}$ is a unit
vector whose $u$th entry equals to one and all other entries equal
to zeros, and ${{\boldsymbol{\tilde{A}}}_{a}}\triangleq \left[
   {{{\boldsymbol{\tilde{a}}}}_{a,1}} \phantom{0}
\cdots \phantom{0} {{{\boldsymbol{\tilde{a}}}}_{a,U}}  \right]$.
Similarly, we have
\begin{small}
\begin{align}
\frac{\partial L(\boldsymbol{p})}{\partial {{\xi
}_{u}}}=2\operatorname{Re}\left\{
\boldsymbol{e}_{u}^{T}\frac{1}{{{\sigma }^{2}}}{{\left(
\boldsymbol{C}\odot \boldsymbol{B} \right)}^{T}}{{\left(
\boldsymbol{Y}_{\left( 1 \right)}^{T}-\left( \boldsymbol{C}\odot
\boldsymbol{B} \right){{\boldsymbol{A}}^{T}}
\right)}^{*}}{{{\boldsymbol{\tilde{A}}}}_{e}}{{\boldsymbol{e}}_{u}}
\right\}
\end{align}
\end{small}
where $ {{\boldsymbol{\tilde{A}}}_{e}}\triangleq \left[
   {{{\boldsymbol{\tilde{a}}}}_{e,1}} \phantom{0}
\cdots \phantom{0} {{{\boldsymbol{\tilde{a}}}}_{e,U}}\right] $,
${{\boldsymbol{\tilde{a}}}_{e,u}}=j{{\boldsymbol{V}}^{T}}
{{\boldsymbol{D}}_{2}}{{\boldsymbol{a}}_{\text{IRS}}}\left(
{{\zeta}_{u}},{{\xi }_{u}} \right)$, and
\begin{align}
\boldsymbol{D}_{2} \triangleq \operatorname{diag}(\underbrace{0,
\cdots,M_{\rm{z}}-1}_{M_{\rm{y}}}, \underbrace{0,\cdots,
M_{\rm{z}}-1}_{M_{\rm{y}}}, \cdots, \underbrace{0, \cdots,
M_{\rm{z}}-1}_{M_{\rm{y}}})
\end{align}
The calculations of $\frac{\partial L(\boldsymbol{p})}{\partial
{{\phi }_{u}}}, \frac{\partial L(\boldsymbol{p})}{\partial
{{\theta }_{u}}}, \frac{\partial L(\boldsymbol{p})}{\partial
{{\varrho }_{u}}}$ and $\frac{\partial L(\boldsymbol{p})}{\partial
{{\iota }_{u}}}$ are similar, which is omitted here.

\subsection{Calculation of Fisher Information Matrix $\Omega\left( \boldsymbol{p} \right)$}
We first calculate the entries in the principal minors of $\Omega
\left( \boldsymbol{p} \right)$. For instance, the $(
{{u}_{1}},{{u}_{2}})$th entry of $\mathbb{E}\{ {{( \frac{\partial
L(\boldsymbol{p})}{\partial {{\boldsymbol{\zeta }}}} )}^{H}}(
\frac{\partial L(\boldsymbol{p})}{\partial {{\boldsymbol{\zeta
}}}})\}$ is given by
\begin{align}
& \mathbb{E}\left\{ {{\left( \frac{\partial
L(\boldsymbol{p})}{\partial {{\zeta }_{{{u}_{1}}}}}
\right)}^{H}}\left( \frac{\partial L(\boldsymbol{p})}{\partial {{\zeta }_{{{u}_{2}}}}}
\right) \right\}\nonumber\\
& =4\mathbb{E}\left\{ \operatorname{Re}\left\{ \boldsymbol{e}_{u}^{T}
{{\boldsymbol{W}}_{1}}{{\boldsymbol{e}}_{u}} \right\}\operatorname{Re}
\left\{ \boldsymbol{e}_{u}^{T}{{\boldsymbol{W}}_{1}}{{\boldsymbol{e}}_{u}} \right\} \right\} \nonumber\\
& =\mathbb{E}\left\{ \left( {{\boldsymbol{W}}_{1}}\left( {{u}_{1}},{{u}_{1}}
\right)+{{\boldsymbol{W}}_{1}}{{\left( {{u}_{1}},{{u}_{1}} \right)}^{*}} \right)\right.\nonumber\\
 &\ \ \ \ \ \ \ \ \ \ \ \ \ \ \ \ \ \ \left.\left( {{\boldsymbol{W}}_{1}}
\left( {{u}_{2}},{{u}_{2}} \right)+{{\boldsymbol{W}}_{1}}{{\left(
{{u}_{2}},{{u}_{2}} \right)}^{*}} \right) \right\}
\end{align}
where ${{\boldsymbol{W}}_{1}} \triangleq \frac{1}{{{\sigma
}^{2}}}{{(\boldsymbol{C}\odot \boldsymbol{B})}^{T}}{{(
\boldsymbol{Y}_{(1)}^{T}-( \boldsymbol{C}\odot \boldsymbol{B}
){{\boldsymbol{A}}^{T}})}^{*}}{{{\boldsymbol{\tilde{A}}}}_{a}}
\triangleq\frac{1}{{{\sigma }^{2}}}{{(\boldsymbol{C}\odot
\boldsymbol{B})}^{T}}{{(\boldsymbol{N}_{( 1
)}^{H})}}{{{\boldsymbol{\tilde{A}}}}_{a}}$. Let
${{\boldsymbol{w}}_{1}}=\text{vec}({{\boldsymbol{W}}_{1}})$. We
have
\begin{align}
{{\boldsymbol{w}}_{1}}=\frac{1}{{{\sigma }^{2}}}\left(
\boldsymbol{\tilde{A}}_{a}^{T}\otimes {{\left( \boldsymbol{C}\odot
\boldsymbol{B} \right)}^{T}} \right)\text{vec}\left(
\boldsymbol{N}_{\left( 1 \right)}^{H} \right)
\end{align}
where ${{\boldsymbol{N}}_{(1)}}$ is the mode-1 unfolding of
$\boldsymbol{\cal{N}}$, $\text{vec}( \boldsymbol{N}_{(1)}^{H})\sim
\mathcal{C}\mathcal{N}(0,{{\sigma }^{2}}\boldsymbol I)$. Since
${{\boldsymbol{w}}_{1}}$ is a linear transformation of
$\text{vec}(\boldsymbol{N}_{(1)}^{H})$, it also follows a
circularly symmetric complex Gaussian distribution. Its covariance
matrix
${{\boldsymbol{C}}_{{{\boldsymbol{w}}_{1}}}}\in\mathbb{C}^{U^2\times
U^2}$ and  second-order moments
${{\boldsymbol{M}}_{{{\boldsymbol{w}}_{1}}}}\in\mathbb{C}^{U^2\times
U^2}$ are respectively given by
\begin{align}
{{\boldsymbol{C}}_{{{\boldsymbol{w}}_{1}}}}& =\mathbb{E}\left\{
{{\boldsymbol{w}}_{1}}\boldsymbol{w}_{1}^{H} \right\} \nonumber\\
& ={{\left( \frac{1}{{{\sigma }^{2}}} \right)}^{2}}\left(
\boldsymbol{\tilde{A}}_{a}^{T}\otimes {{\left( \boldsymbol{C}\odot
\boldsymbol{B} \right)}^{T}} \right)\nonumber\\
&\ \times \mathbb{E}\left\{ \text{vec}\left( \boldsymbol{N}_{\left( 1
\right)}^{H} \right)\text{vec}{{\left(\boldsymbol{N}_{\left( 1 \right)}^{H}
\right)}^{H}} \right\}{{\left( \boldsymbol{\tilde{A}}_{a}^{T}\otimes {{
\left( \boldsymbol{C}\odot \boldsymbol{B} \right)}^{T}} \right)}^{H}} \nonumber\\
& =\frac{1}{{{\sigma }^{2}}}\left(
\boldsymbol{\tilde{A}}_{a}^{T}\boldsymbol{\tilde{A}}_{a}^{*}
\right)\otimes \left( {{\left( \boldsymbol{C}\odot \boldsymbol{B}
\right)}^{T}}{{\left(\boldsymbol{C}\odot \boldsymbol{B}
\right)}^{*}} \right)
\end{align}
and
\begin{align}
{{\boldsymbol{M}}_{{{\boldsymbol{w}}_{1}}}}=\mathbb{E}\left\{
{{\boldsymbol{w}}_{1}}\boldsymbol{w}_{1}^{T} \right\}=0
\end{align}
Thus we have
\begin{align}
\mathbb{E}\bigg\{ {{\bigg( \frac{\partial
L(\boldsymbol{p})}{\partial {{\zeta}_{{{u}_{1}}}}}
\bigg)}^{H}}\bigg( \frac{\partial L(\boldsymbol{p})}{\partial
{{\zeta}_{{{u}_{2}}}}}\bigg)\bigg\}=2\operatorname{Re}\left\{
{{\boldsymbol{C}}_{{{\boldsymbol{w}}_{1}}}}\left( m,n \right)
\right\}
\end{align}
where $m\triangleq U\left( {{u}_{1}}-1
\right)+{{u}_{1}},n\triangleq U\left( {{u}_{2}}-1
\right)+{{u}_{2}}$. Similarly, we have
\begin{align}
\mathbb{E}\bigg\{ {{\bigg( \frac{\partial
L(\boldsymbol{p})}{\partial {{\xi}_{{{u}_{1}}}}}
\bigg)}^{H}}\bigg( \frac{\partial L(\boldsymbol{p})}{\partial
{{\xi}_{{{u}_{2}}}}} \bigg) \bigg\}&=2\operatorname{Re}\left\{
{{\boldsymbol{C}}_{{{\boldsymbol{w}}_{2}}}}\left( m,n \right)
\right\}
\end{align}
where
${{\boldsymbol{C}}_{{{\boldsymbol{w}}_{2}}}}=\frac{1}{{{\sigma
}^{2}}}(
\boldsymbol{\tilde{A}}_{e}^{T}\boldsymbol{\tilde{A}}_{e}^{*}
)\otimes ( {{(\boldsymbol{C}\odot \boldsymbol{B} )}^{T}}{{(
\boldsymbol{C}\odot \boldsymbol{B})}^{*}})$. The derivations of
other entries in the principal minors of $\Omega \left(
\boldsymbol{p} \right)$ are similar and thus omitted here.

For elements in the off-principal minors of $\Omega(
\boldsymbol{p})$, such as the $(l_1, l_2)$th entry of
$\mathbb{E}\{ {{(\frac{\partial L(\boldsymbol{p})}{\partial
{{\boldsymbol{\zeta }}}})}^{H}}(\frac{\partial
L(\boldsymbol{p})}{\partial {{\boldsymbol{\xi }}}})\}$, we have
\begin{align}
  \mathbb{E}\bigg\{ {{\bigg( \frac{\partial L(\boldsymbol{p})}
  {\partial {{\zeta }_{{{u}_{1}}}}} \bigg)}^{H}}\bigg(
  \frac{\partial L(\boldsymbol{p})}{\partial {{\xi }_{{{u}_{2}}}}}
  \bigg) \bigg\}=2\operatorname{Re}\left\{ {{\boldsymbol{C}}_{{{\boldsymbol{w}}_{1}},
  {{\boldsymbol{w}}_{2}}}}\left( m,n \right) \right\}
\end{align}
where
\begin{align}
{{\boldsymbol{C}}_{{{\boldsymbol{w}}_{1}},{{\boldsymbol{w}}_{2}}}}&=
\mathbb{E}\left\{ {{\boldsymbol{w}}_{1}}\boldsymbol{w}_{2}^{H} \right\} \nonumber\\
& ={{\left( \frac{1}{{{\sigma }^{2}}} \right)}^{2}}\left(
\boldsymbol{\tilde{A}}_{a}^{T}\otimes {{\left(
\boldsymbol{C}\odot \boldsymbol{B} \right)}^{T}} \right)\mathbb{E}
\left\{ \text{vec}\left(\boldsymbol{N}_{\left( 1 \right)}^{H} \right)\right.\nonumber\\
 &\left.\ \ \ \ \ \ \ \ \ \ \ \times\text{vec}{{\left( \boldsymbol{N}_{
 \left( 1 \right)}^{H} \right)}^{H}} \right\}{{\left(
 \boldsymbol{\tilde{A}}_{b}^{T}\otimes {{\left( \boldsymbol{C}\odot \boldsymbol{B}
 \right)}^{T}} \right)}^{H}} \nonumber\\
 & =\frac{1}{{{\sigma }^{2}}}\left( \boldsymbol{\tilde{A}}_{a}^{T}\boldsymbol{\tilde{A}}_{b}^{*}
 \right)\otimes \left( {{\left(\boldsymbol{C}\odot \boldsymbol{B} \right)}^{T}}{{
 \left(\boldsymbol{C}\odot \boldsymbol{B} \right)}^{*}} \right)
\end{align}
Other entries in the off-principal minors of $\Omega \left(
\boldsymbol{p} \right)$ can be similarly calculated.

\subsection{Cram\'{e}r-Rao Bound}
After obtaining the FIM, the CRB for the parameters $\boldsymbol
p$ can be calculated as
\begin{align}
 CRB \left( \boldsymbol p \right)= \Omega^{-1} \left( {\boldsymbol{p}} \right)
\end{align}

\bibliography{newbib}
\bibliographystyle{IEEEtran}

\end{document}